\documentclass[11pt,a4paper]{article}

\usepackage{amsmath,amssymb,bm}
\usepackage{color}
\usepackage{multirow}

\usepackage{graphicx}
\usepackage{hyperref}

\usepackage{tikz}

\newcommand{\Z}{\mathbb{Z}}
\newcommand{\C}{\mathbb{C}}

\newcommand{\N}{\mathcal{N}}

\newcommand{\cC}{\mathcal{C}}

\newcommand{\cK}{\mathcal{K}}
\newcommand{\cO}{\mathcal{O}}
\newcommand{\cS}{\mathcal{S}}
\newcommand{\cZ}{\mathcal{Z}}

\newcommand{\U}{\mathrm{U}}
\newcommand{\bbP}{\mathbb{P}}
\newcommand{\SU}{\mathrm{SU}}
\newcommand{\SO}{\mathrm{SO}}
\newcommand{\Sp}{\mathrm{Sp}}
\newcommand{\SL}{\mathrm{SL}}

\newcommand{\Tr}{\operatorname{Tr}}

\newcommand{\Str}{\operatorname{Str}}
\newcommand{\Sdet}{\operatorname{Sdet}}

\newcommand{\unknot}{{\,\raisebox{-.08cm}{\includegraphics[width=.37cm]{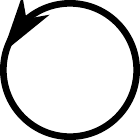}}\,}}

\makeatletter

\@addtoreset{equation}{section}
\makeatother

\date{\today}

\begin{document}

\begin{titlepage}

\renewcommand{\thefootnote}{\fnsymbol{footnote}}

\begin{flushright}
 {\tt 
 IPHT-T14/088 \\
 CRM-3338 \\
 RIKEN-MP-90
 }
\\
\end{flushright}

\vskip8em

\begin{center}
 {\Large {\bf 
 Towards $\U(N|M)$ knot invariant from ABJM theory
 }}

 \vskip5em

 \setcounter{footnote}{1}
 {\sc Bertrand Eynard$^{a,b}$}\footnote{E-mail address: 
 \href{mailto:bertrand.eynard@cea.fr}
 {\tt bertrand.eynard@cea.fr}} 
 and
 \setcounter{footnote}{2}
 {\sc Taro Kimura$^{a,c}$}\footnote{
 Present address: Department of Physics, Keio University;
 E-mail address: 
 \href{mailto:taro.kimura@keio.jp}
 {\tt taro.kimura@keio.jp}\\
 \qquad
 Keywords:
 knot invariant, ABJM theory, Chern--Simons theory, Matrix model\\
 MSC: 81T60, 81T45, 57M27, 14H81
 }

 \vskip2em

{\it 
 $^{a}$Institut de Physique Th\'eorique,
 CEA Saclay, F-91191 Gif-sur-Yvette, France
 \\ \vspace{.5em}
 $^{b}$Centre de Recherches Math\'ematiques, 
 Universit\'e de Montr\'eal,
 Montr\'eal, QC, Canada,
 \\ \vspace{.5em}
 $^{c}$Mathematical Physics Laboratory, RIKEN Nishina Center, 
 Saitama 351-0198, Japan 
}

 \vskip3em

\end{center}

 \vskip2em

\begin{abstract}
We study $\U(N|M)$ character expectation value with the supermatrix Chern--Simons theory, known as the ABJM matrix model, with emphasis on its connection to the knot invariant.
This average just gives the half-BPS circular Wilson loop expectation value in ABJM theory, which shall correspond to the unknot invariant.
We derive the determinantal formula, which gives $\U(N|M)$ character expectation values in terms of $\U(1|1)$ averages for a particular type of character representations. 
This means that the $\U(1|1)$ character expectation value is a building block for the $\U(N|M)$ averages, and also, by an appropriate limit, for the $\U(N)$ invariants.
In addition to the original model, we introduce another supermatrix model obtained through the symplectic transform, which is motivated by the torus knot Chern--Simons matrix model.
We obtain the Rosso--Jones-type formula and the spectral curve for this case.

\end{abstract}

\end{titlepage}

\tableofcontents

\hrulefill

\setcounter{footnote}{0}


\section{Introduction}\label{sec:intro}

Since it was shown by Witten~\cite{Witten:1988hf} that a knot
invariant is realized by using the Wilson loop operators in
Chern--Simons gauge theory, knot theory has been providing various kinds
of interesting topics not only for mathematicians, but also for physicists.
In particular the most important example of the knot invariant, which is
called the Jones polynomial, is obtained from Chern--Simons theory with the
Wilson loop in a fundamental representation
\begin{equation}
 J(K;q) 
  = 
  \Big\langle
  W_\square (K;q)
  \Big\rangle
  \Big /
  \Big\langle
  W_{\square} (\unknot;q)
  \Big\rangle
  \, ,
  \label{WL_normalize}
\end{equation}
where the Wilson loop in a representation $R$ is given by
\begin{equation}
 W_R(K;q) 
  =
  \mathrm{Tr}_R \, \mathrm{P}
  \exp \left( \oint_K A \right)
  \, ,
\end{equation}
and its expectation value is taken with respect to Chern--Simons theory
with $\SU(2)$ gauge symmetry on a three-sphere $S^3$
\begin{equation}
 S_{\rm CS}[A]
  =
  \frac{k}{4\pi} \int_{S^3} \Tr 
  \left( A \wedge dA + \frac{2}{3} A \wedge A \wedge A \right)
  \, .
\end{equation}
In this case the parameter $q$ is associated with the level of
Chern--Simons theory as $q = \exp \left( 2\pi i / (k+2) \right)$.
This prescription to derive the knot invariant is quite general:
when the fundamental representation $R = \square$ is replaced by a generic
representation, one obtains the colored Jones polynomial, and $\SU(N)$
and $\SO(N)/\Sp(N)$ generalizations provide the HOMFLY and Kauffman
polynomials, respectively.

The knot polynomial is usually defined by the Skein relation with a
proper normalization condition for the unknot invariant.
Although it is in principle computable for any knots based on that
definition, their expressions get much complicated as the number of 
crossings in knots increases, or the representation of the knot
polynomial becomes highly involved.
Thus it is not yet achieved to write down such a knot invariant in a
closed form.
On the other hand, for a particular class of knots, the unknot and also
torus knots, there is a useful integral representation of the knot
invariant~\cite{Lawrence:1999CMP,Beasley:2009mb,Kallen:2011ny,Brini:2011wi},
which is applicable to a generic gauge group and representation.
This is based on the matrix integral formula for the partition function
of Chern--Simons theory, especially defined on a three-sphere $S^3$, and
then given as the expectation value of the character in the
corresponding group.

In this paper we consider a supergroup character average with the supermatrix
Chern--Simons theory, known as the ABJM matrix
model~\cite{Kapustin:2009kz}, towards a supersymmetric generalization of
the knot invariant.
This supermatrix model is derived from $\N=6$ superconformal
Chern--Simons--matter theory with gauge group $\U(N)_k \times
\U(N)_{-k}$, which is so-called ABJM theory~\cite{Aharony:2008ug}, by
implementing the localization technique for the path integral.
In this theory the circular Wilson loop, in particular the half-BPS
operator, is described by a holonomy with a superconnection taking a
value in $\U(N|N)$, which is written as a supergroup
character~\cite{Drukker:2009hy}.
Thus it is expressed in terms of the
supersymmetric Schur function~\cite{Bars:1982ps,Berele:1987yi}.
In this sense the average we compute here is expected to be an unknot
invariant for $\U(N|N)$ theory.

The knot invariant and the character average have an
analogous structure to a matrix integral in the presence of an
external field, which is dual to a correlation function of
characteristic polynomials.
Especially in the case of supermatrix models, the corresponding correlation
function is that for the characteristic polynomial ratio.
For this correlation function, there is an interesting determinantal
formula which consists of a single pair correlation function as a
kernel~\cite{Fay:1973,Borodin:2006AAM,Bergere:2009zm}.
In this way we shall expect that a similar determinantal structure can
be found in $\U(N|N)$ theory, and the $\U(1|1)$
expectation value plays a role as the kernel function there.
We will show in this paper such a determinantal formula for the 
average with a particular type of the character representation.

According to the general scheme of the topological
recursion~\cite{Eynard:2007kz}, one can determine all order perturbation
series for the correlation function from the spectral curve.
In the case of the knot invariant, there are basically two kinds of
perturbative expansions.
The first is based on the spectral curve, which is obtained from the
A-polynomial, and its expansion is from the large representation
limit~\cite{Dijkgraaf:2010ur,Borot:2012cw}.
Although this A-polynomial was originally introduced to the Jones
polynomial, namely $\SU(2)$ Chern--Simons theory, it can be now extended to
more generic theories.
See for example~\cite{Gukov:2012jx}.
The other expansion comes from the spectral curve arising in the large
rank limit of the knot invariant.
This kind of spectral curves is quite analogous to that discussed in
random matrix theory, because for some kinds of knots we have matrix
integral-like expressions for the knot invariant, and in this case the
matrix size $N$ corresponds to the rank of Chern--Simons gauge group
$\SU(N)$.
This large $N$ limit plays an important role in topological string
theory, because it describes the geometric transition of the
corresponding Calabi--Yau threefold~\cite{Gopakumar:1998ki}.
Such a duality is still available for the situation in the presence
of a knot~\cite{Ooguri:1999bv}, which gives a brane on a proper
Lagrangian submanifold of the Calabi--Yau threefold.
In the sense of topological strings, the corresponding spectral curve
provides the mirror Calabi--Yau threefold.
In this paper we will discuss the large $N$ spectral curve for the
supermatrix Chern--Simons theories.

The supergroup character average discussed in this paper is based on
ABJM theory.
As pointed out in~\cite{Marino:2009jd}, it is perturbatively equivalent to
Chern--Simons theory on the lens space $L(2,1) = S^3/\Z_2 \cong
\mathbb{RP}^3$, which is dual to the topological string on the local
$\mathbb{P}^1 \times \mathbb{P}^1$ geometry, and the associated spectral
curve becomes a genus-one curve.
Therefore the unknot spectral curve is just given by the mirror curve of
the local $\mathbb{P}^1 \times \mathbb{P}^1$.
In the case of the torus knot, the spectral curve is obtained by
applying the symplectic transformation of the
unknot~\cite{Brini:2011wi}.
As well as the ordinary HOMFLY polynomial for $\SU(N)$ Chern--Simons
theory, we will introduce the $(P,Q)$-deformed supermatrix model through
the symplectic transform of the original matrix model, which
is motivated by the torus knot Chern--Simons matrix model.
Since the Adams operation works well even for the Schur function
associated with the supergroup $\U(N|N)$, we can derive the Rosso--Jones
formula for the torus knot average.
This means that the torus knot character average can be represented as a linear
combination of the fractionally framed unknot averages.
We will also derive the spectral curve from the saddle point equations
of the $(P,Q)$-deformed supermatrix model, and then obtain a consistent result
with the symplectic transform of the unknot curve.

This paper is organized as follows.
In Sec.~\ref{sec:unknot} we study the supergroup character average with
the ABJM matrix model, which is just given as the half-BPS circular
Wilson loop operator in ABJM theory.
We especially focus on the representation, corresponding to the
partition such that the number of its diagonal components is given by $N$.
We will show the determinantal formula factorizing the
$\U(N|N)$ character average into the $\U(1|1)$ expectation values.
In Sec.~\ref{sec:torus} we then consider the $(P,Q)$-deformation of the
supermatrix model, which is obtained by the
$\mathrm{SL}(2,\Z)$ transform of the original ABJM matrix model.
We will show that the Rosso--Jones formula holds even for the
$\U(N|N)$ theoy, and the $\U(1|1)$ average plays a role of a
building block for the $(P,Q)$-deformed $\U(N|N)$ character expectation value
with a particular kind of representations.
In Sec.~\ref{sec:U(N|M)} we extend the argument to $\U(N|M)$ theory, and
we will see that the determinantal formula for the character average can be
derived even in this case.
We obtain the expression which interpolates $\U(N)$ and $\U(N|N)$ theories.
In Sec.~\ref{sec:sp_curve} we discuss the spectral curve for the
$(P,Q)$-deformed matrix model.
We show that there are two consistent ways of obtaining the spectral
curve: One is the symplectic transform of the original matrix model and the
other is the saddle point analysis of the $(P,Q)$-deformed matrix model
itself.
In Sec.~\ref{sec:comments} we give some comments on the relations
to topological string and random matrix theory.
Sec.~\ref{sec:summary} is devoted to a summary and discussions.

\section{Unknot matrix model}\label{sec:unknot}

Let us start with the matrix model description of Chern--Simons theory
at level $k$ defined on a three-sphere $S^3$.
When we take the gauge group $G = \U(N)$, the partition function is
given as the matrix model-like integral
\begin{equation}
 \cZ_{\rm CS} (S^3;q)
  = 
  \frac{1}{N!} \int \! \prod_{i=1}^N \frac{dx_i}{2\pi} \, 
  e^{-\frac{1}{2g_s} x_i^2} \,
  \prod_{i<j}^N
  \left(
   2 \sinh \frac{x_i-x_j}{2}
  \right)^2
  \, ,
  \label{CS_MM}
\end{equation}
where the parameter $q$ is related to the coupling constant as $q =
\exp g_s$ with
\begin{equation}
 g_s = \frac{2\pi i}{k+N}
  \, .
\end{equation}
We then consider the expectation value of the circular Wilson loop
operator with representation $R$, which corresponds to the unknot
invariant, 
\begin{equation}
 \Big\langle
  W_R(\unknot)
 \Big\rangle
 =
 \frac{1}{\cZ_{\rm CS}} \frac{1}{N!} 
 \int \! \prod_{i=1}^N \frac{dx_i}{2\pi} \, 
  e^{-\frac{1}{2g_s} x_i^2} \,
  \prod_{i<j}^N
  \left(
   2 \sinh \frac{x_i-x_j}{2}
  \right)^2 \,
  \mathrm{Tr}_R \, U(x)
  \, .
  \label{CS_WL}
\end{equation}
The holonomy matrix $U(x)$ is given by 
\begin{equation}
 U(x) = \left(
      \begin{array}{ccc}
       e^{x_1} & & \\
        & \ddots & \\
        & & e^{x_N}
      \end{array}
     \right)
 \, ,
\end{equation}
and $\mathrm{Tr}_R \, U$ is indeed the character of $G=\U(N)$ in the
representation $R$.
This character is written as a Schur polynomial with the
corresponding partition $\lambda$ to the representation $R$
\begin{equation}
 \mathrm{Tr}_R \, U = s_\lambda (e^{x_1}, \cdots, e^{x_N})
  \, .
\end{equation}
Although we can deal with only the unknot Wilson loop based on this matrix
model, it is possible to obtain the torus knot invariants by applying
a slightly different matrix model, as discussed in Sec.~\ref{sec:torus}.

We now consider a supersymmetric extension of this Chern--Simons theory.
In this paper we especially apply the supermatrix generalization of the
Chern--Simons matrix model (\ref{CS_MM}), namely the ABJM matrix
model~\cite{Drukker:2009hy}. 
It was shown in \cite{Kapustin:2009kz} that ABJM
theory~\cite{Aharony:2008ug}, which is the three-dimensional
superconformal Chern--Simons--matter theory with gauge group $\U(N)_k
\times \U(N)_{-k}$, can be similarly reduced to the matrix model-like integral
\begin{eqnarray}
 \cZ_{\rm ABJM} (S^3;q)
  & = &
  \frac{1}{N!^2} \int \! \prod_{i=1}^N \frac{dx_i}{2\pi} \frac{dy_i}{2\pi} \, 
  e^{-\frac{1}{2g_s} \left( x_i^2 - y_i^2 \right)} \,
  \prod_{i,j}^N
  \left(
   2 \cosh \frac{x_i-y_j}{2}
  \right)^{-2}
  \nonumber \\
 & & 
  \hspace{8em}
  \times
  \prod_{i<j}^N
  \left(
   2 \sinh \frac{x_i-x_j}{2}
  \right)^2
  \left(
   2 \sinh \frac{y_i-y_j}{2}
  \right)^2
  \, .
  \label{ABJM_MM}
\end{eqnarray}
In this case there is no level shift in the coupling constant
\begin{equation}
 g_s = \frac{2\pi i}{k}
  \, .
\end{equation}
We can insert the Wilson loop operator into this matrix model as well as
Chern--Simons theory with the classical group.
Although there are some possibilities for the operators in this case,
the relevant choice to this study is the half-BPS circular Wilson loop
operator, which is given by a character of the supergroup
$\U(N|N)$~\cite{Drukker:2009hy,Marino:2009jd},
\begin{eqnarray}
 \Big\langle
  W_R(\unknot)
 \Big\rangle
 & = & 
 \frac{1}{\cZ_{\rm ABJM}} \frac{1}{N!^2}
 \int \! \prod_{i=1}^N \frac{dx_i}{2\pi} \frac{dy_i}{2\pi} \, 
  e^{-\frac{1}{2g_s} \left( x_i^2 - y_i^2 \right)} \,
  \prod_{i,j}^N
  \left(
   2 \cosh \frac{x_i-y_j}{2}
  \right)^{-2}
  \nonumber \\
 & & 
  \hspace{4em}
  \times
  \prod_{i<j}^N
  \left(
   2 \sinh \frac{x_i-x_j}{2}
  \right)^2
  \left(
   2 \sinh \frac{y_i-y_j}{2}
  \right)^2
  \, 
  \mathrm{Str}_R \, U(x;y)
  \, ,
  \label{ABJM_WL}
\end{eqnarray}
where the matrix $U(x;y)$ is of the size $2N \times 2N$
\begin{equation}
 U(x;y) 
  = \left(
     \begin{array}{cc}
      U(x) & \\
      & -U(y) \\
     \end{array}
    \right)
 \, .
\end{equation}
Let us call this expectation value the unknot Wilson loop average as an
analogy with the knot invariant.
The supergroup character for $\U(N|N)$ is obtained by replacing the
power sum polynomial $\Tr U^n$ in the $\U(N+N)$ character 
with the supertrace $\Str U^n$~\cite{Bars:1982ps}.

As well as the $\U(N)$ representation theory, the supergroup character
can be also expressed as the Schur
polynomial, but with a prescribed symmetry~\cite{Berele:1987yi}
\begin{equation}
 \mathrm{Str}_R \, U(x;y)
  =
  s_\lambda (e^x;e^y)
  \, .
\end{equation}
For $\U(N|N)$ theory, we have a useful determinantal
formula in terms of the Frobenius coordinate of the
partition $\lambda=(\alpha_1, \cdots, \alpha_{d(\lambda)}|\beta_1, \cdots,
\beta_{d(\lambda)})$ with $\alpha_i = \lambda_i - i$ and $\beta_i =
\lambda^t_i - i$~\cite{Moens:2003JAC}
\begin{equation}
 s_\lambda(u;v)
  =
  \det_{1 \le i, j \le d(\lambda)}
  \left(
   \sum_{k,l=1}^N 
   u_k^{\alpha_i}
   \left( C^{-1} \right)_{kl}
   v_l^{\beta_j}
  \right)
  \, ,
  \label{super_Schur01}
\end{equation}
where the matrix $C^{-1}$ is the inverse of the Cauchy matrix
\begin{equation}
 C = 
  \left(
   \frac{1}{u_k+v_l}
  \right)_{1 \le k, l \le N}
  \, .
\end{equation}
We remark that the supersymmetric Schur polynomial is identically zero
when $d(\lambda) > N$, or equivalently $\lambda_{N+1} > N$.

The formula (\ref{super_Schur01}) also implies that it can be written only in
terms of the hook representations
\begin{equation}
 s_{\lambda} (u;v)
  = 
  \det_{1 \le i, j \le d(\lambda)} 
  s_{(\alpha_i|\beta_j)} (u;v)
  \, .
\end{equation}
This is just a supersymmetric version of the Giambelli formula.
Actually this relation is useful to study the Wilson loop
operators with various representations in ABJM theory~\cite{Hatsuda:2013yua}.

\begin{figure}[t]
 \begin{center}
  \begin{tikzpicture}
   \path (0,0) node {$\lambda \ =$};
   \path (1.75,0.75) node {$N \times N$};
   \path (4.15,-0.05) node {$\mu$};
   \path (2.75,-1.35) node {$\nu^t$};

   \draw 
   (1,0) -- (1,1.5) -- (2.5,1.5) -- (2.5,0) -- (1,0) ;

   \filldraw [fill=gray,opacity=.6,draw=black,thick]
   (2.5,1.5) -- (4.6,1.5) -- (4.6,1.2) -- (3.7,1.2) -- (3.7,0.9)
   -- (3.4,0.9) -- (3.4,0.6) -- (2.8,0.6) -- (2.8,0.3) 
   -- (2.5,0.3) -- (2.5,1.5);

   \filldraw [fill=gray,opacity=.6,draw=black,thick]
   (1,0) -- (1,-1.5) -- (1.6,-1.5) -- (1.6,-0.9) -- (1.9,-0.9)
   -- (1.9,-0.6) -- (2.2,-0.6) -- (2.2,-0.3) -- (2.5,-0.3) 
   -- (2.5,0) -- (1,0);

   \draw [<-,thick] 
   (3.2,0.9) .. controls (3.4,0.3) and (3.5,0.3) .. (3.65,0.6) 
   .. controls (3.8,0.8) and (3.9,0.8) .. (4.1,0.2) ;

   \draw [<-,thick]
   (1.75,-0.7) .. controls (2.4,-0.8) and (2.4,-0.9) .. (2.1,-1.1)
   .. controls (2,-1.2) and (2,-1.3) .. (2.5,-1.4);

  \end{tikzpicture}
 \end{center}
 \caption{The partition $\lambda = (12,9,8,6,5,5,4,3,2,2)$, which is
 also represented as $\lambda = (11,7,5,2,0|9,8,5,3,1)$ in the Frobenius
 coordinate with $d(\lambda) = N \, (=5)$.
 We obtain sub diagrams $\mu=(7,4,3,1,0)$ and  $\nu^t=(5,4,3,2,2)$
 involved in this partition.
 }
 \label{fig:partition}
\end{figure}
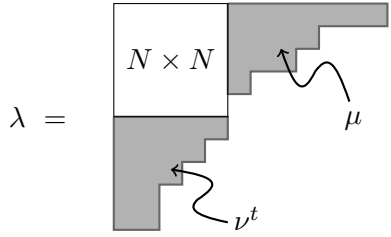

Especially for the most generic situation with $d(\lambda)=N$, on which
we focus in this paper, the supersymmetric Schur polynomial is
decomposed into the ordinary ones~\cite{Moens:2003JAC}
\begin{eqnarray}
 s_\lambda(u;v)
  & = &
  \det_{1 \le i, j \le N} u_i^{\alpha_j} \, \cdot
  \det_{1 \le i, j \le N} v_i^{\beta_j} \,  \cdot
  \det_{1 \le i, j \le N} C_{ij}^{-1}
  \nonumber \\
 & = &
  s_{\mu}(u) \, s_\nu (v) \,
  \prod_{i,j=1}^N (u_i + v_j)
  \, ,
  \label{Schur_factorization}
\end{eqnarray}
where the partitions $\mu$ and $\nu$ are given by
\begin{equation}
 \mu_i = \lambda_i - N \, , \qquad
 \nu_i = \lambda^t_i - N \, , \qquad
 i = 1, \cdots, N \, ,
\end{equation}
or equivalently, 
$\mu^t_i = \lambda^t_{i+N}$, $\nu^t_i=\lambda_{i+N}$,
as shown in Fig.~\ref{fig:partition}. 
The determinant of the Cauchy matrix is given by
\begin{equation}
 \det C 
  = 
  \Delta(u) \Delta(v)
  \prod_{i,j=1}^N (u_i + v_j)^{-1}
  \, ,
  \label{Cauchy_det}
\end{equation}
with the Vandermonde determinant
\begin{equation}
 \Delta(u) = \prod_{i<j}^N (u_i - u_j)
  \, .
\end{equation}

\subsection{\texorpdfstring{$\U(1|1)$}{U(1|1)} theory}

Let us consider the simplest example with the supergroup $\U(1|1)$,
which plays a fundamental role in this study.
In this case only the hook representation is possible, which is written
as $\lambda = (\alpha|\beta)$ with the Frobenius coordinate.
The corresponding Schur function reads
\begin{equation}
 s_{(\alpha|\beta)} (e^x;e^y)
  =
  \left(
   e^x + e^y
  \right)
  \,
  e^{\alpha x + \beta y}
  \, .
\end{equation}
The unknot Wilson loop expectation value is given by
\begin{eqnarray}
 \Big\langle
  W_{(\alpha|\beta)}(\unknot)
 \Big\rangle
 =
 \frac{1}{\cZ_{\rm ABJM}}
 \int \frac{dx}{2\pi} \frac{dy}{2\pi} \,
 e^{\frac{ik}{4\pi}(x^2-y^2)}
 \left(
  2 \cosh \frac{x-y}{2}
 \right)^{-2}
  \left(
   e^x + e^y
  \right)
 e^{\alpha x + \beta y}
 \, .
\end{eqnarray}
We can compute this integral explicitly by applying the 
Fourier transform formula
\begin{equation}
 \frac{1}{2 \cosh w}
  =
  \int \! \frac{dz}{2\pi} \,
  \frac{e^{2i w z / \pi}}{\cosh z}
  \, .
  \label{cosh_FT}
\end{equation}
Thus we have
\begin{align}
 \int \frac{dx}{2\pi} \frac{dy}{2\pi} \frac{dz}{2\pi} \,
  \frac{1}{\cosh z} \, 
 e^{\frac{ik}{4\pi} (x^2-y^2) 
	+ \left( \alpha + \frac{1}{2} \right) x
	+ \left( \beta + \frac{1}{2} \right) y
	+ \frac{i}{\pi} (x-y) z}
 & =
 \frac{1}{k} \,
 \frac{q^{\frac{1}{2}(\alpha+\beta+1)(\alpha-\beta)}}
      {q^{\frac{1}{2}(\alpha+\beta+1)}+q^{-\frac{1}{2}(\alpha+\beta+1)}}
 \, .
\end{align}
Since the partition function becomes $\cZ_{\rm ABJM} = (4k)^{-1}$
for $N=1$, we obtain the expectation value as follows,
\begin{equation}
 \Big\langle
  W_{(\alpha|\beta)}(\unknot)
 \Big\rangle
 = 
 \frac{4 \, q^{\frac{1}{2}(\alpha+\beta+1)(\alpha-\beta)}}
      {q^{\frac{1}{2}(\alpha+\beta+1)}+q^{-\frac{1}{2}(\alpha+\beta+1)}}
 \, .
 \label{WL_U(11)}
\end{equation}
This depends only on the total and relative lengths of the partition,
given by  $\alpha+\beta+1$ and $\alpha-\beta$, respectively.
We will see that the numerator can be seen as the framing factor in the
following section.

\subsection{\texorpdfstring{$\U(N|N)$}{U(N|N)} theory}

We then consider the character expectation value for $\U(N|N)$ theory, in
particular with a representation with $d(\lambda) = N$,
$\lambda=(\alpha_1, \cdots, \alpha_N|\beta_1, \cdots, \beta_N)$.
Let us first rewrite the partition function (\ref{ABJM_MM}) 
\begin{equation}
 \cZ_{\rm ABJM} (S^3;q)
  =
  \frac{1}{N!^2}
  \int [dx]^N [dy]^N \,
  \det
  \left(
   \frac{1}{2 \cosh \frac{x_i-y_j}{2}}
  \right)^2
  \, ,
\end{equation}
with shorthand notations
\begin{equation}
 [dx] = \frac{dx}{2\pi} \, e^{-\frac{1}{g_s} x^2} \, , \qquad
 [dy] = \frac{dy}{2\pi} \, e^{\frac{1}{g_s} y^2} \, ,
\end{equation}
where we have used the Cauchy formula
\begin{equation}
 \det_{1 \le i, j \le N}
 \left(
  \frac{1}{2 \cosh \frac{x_i - y_j}{2}}
 \right)
 =
 \prod_{i,j=1}^N
  \left(
   2 \cosh \frac{x_i-y_j}{2}
  \right)^{-1}
  \prod_{i<j}^N
  \left(
   2 \sinh \frac{x_i-x_j}{2}
  \right)
  \left(
   2 \sinh \frac{y_i-y_j}{2}
  \right)
  \, .
\end{equation}
In this case the Schur function has a simple expression as shown in
(\ref{Schur_factorization}).
Therefore the unnormalized unknot expectation value is now given by
\begin{equation}
 \Big\langle
  W_{R}(\unknot)
 \Big\rangle
 = 
  \int [dx]^N [dy]^N \,
  \det_N
  \left(
   \frac{1}{2 \cosh \frac{x_i-y_j}{2}}
  \right) \,
  \det_N e^{x_i \xi_j}
  \det_N e^{y_i \eta_j}
  \, ,
  \label{ABJM_WL_ext}
\end{equation}
where the parameters $\xi_i$ and $\eta_i$, defined as $\xi_i = \lambda_i
- i + 1/2 = \alpha_i + 1/2$ and $\eta_i = \lambda^t_i - i + 1/2 =
\beta_i + 1/2$, play a similar role to external fields in matrix
models, as discussed in Sec.~\ref{sec:comments}.
Since all the $x_i$ and $y_i$ are not distinguishable,
this integral can be expressed as a size $N$ determinant
\begin{equation}
 \det_{1 \le i, j \le N}
 \left[
  \int \frac{dx}{2\pi} \frac{dy}{2\pi} \,
  e^{\frac{ik}{4\pi}(x^2-y^2) + x \xi_i + y \eta_j}
  \left( 2 \cosh \frac{x-y}{2} \right)^{-1}
 \right]
 \, .
\end{equation}
At this moment the computation is almost reduced to that for $\U(1|1)$ theory.
Again using the formula (\ref{cosh_FT}), we obtain the determinantal
formula for the character expectation value
\begin{equation}
 \Big\langle
  W_{R}(\unknot)
 \Big\rangle
 =
 k^{-N}
 \prod_{i=1}^N
 q^{\frac{1}{2}(\xi_i^2-\eta_i^2)} \,
 \det_{1 \le i, j \le N}
 \left(
  \frac{1}{q^{\frac{1}{2}(\xi_i+\eta_j)}+q^{-\frac{1}{2}(\xi_i+\eta_j)}}
 \right)
 \, .
 \label{unknot_NN01}
\end{equation}
This shows that the $\U(N|N)$ unknot character average is factorized
into that for $\U(1|1)$ theory (\ref{WL_U(11)}), and thus the measure of
this matrix integral is {\em Giambelli compatible} in the sense
of~\cite{Borodin:2006AAM}.

To see that the $\U(N|N)$ theory contains the 
$\U(N)$ knot invariant, it is convenient to rewrite the expression
(\ref{unknot_NN01}) by applying the Cauchy formula,
\begin{align}
 \Big\langle
  W_{R}(\unknot)
 \Big\rangle
 & = 
 k^{-N}
 q^{\frac{1}{2}\left( C_2(\mu) - C_2(\nu)\right)}
 \prod_{i,j=1}^N
 \left(
  q^{\frac{1}{2}(\alpha_i+\beta_j+1)}
  + q^{-\frac{1}{2}(\alpha_i+\beta_j+1)}
 \right)^{-1}
 \nonumber \\
 &  \hspace{3em}
  \times
  \prod_{i<j}^N
  \left(
   q^{\frac{1}{2}(\alpha_i-\alpha_j)} 
   - q^{-\frac{1}{2}(\alpha_i-\alpha_j)}
  \right)
  \left(
   q^{\frac{1}{2}(\beta_i-\beta_j)} 
   - q^{-\frac{1}{2}(\beta_i-\beta_j)}
  \right)
  \, .
  \label{unknot_NN02}
\end{align}
where the 2nd Casimir operator is defined by
\begin{align}
 C_2(\lambda)
 & =
 \sum_{i=1}^\infty \left( \left(\lambda_i - i +
 \frac{1}{2} \right)^2 - \left( - i + \frac{1}{2} \right)^2 \right)
 \, .
\end{align}
Thus in this case we see that the standard framing factor is given by
$q^{\frac{1}{2} (C_2(\mu)-C_2(\nu))}$.
Up to the normalization constants, the factors in the second line
coincide with the Wilson loop expectation value for $\U(N)$ theory,
which is given by the quantum dimension of the representation $R$,
\begin{eqnarray}
 \Big\langle
  W_{R}(\unknot)
 \Big\rangle_{\U(N)}
 & = &
 \prod_{i<j}^N 
 \frac{[\lambda_i-\lambda_j-i+j]_q}{[-i+j]_q}
 \nonumber \\
 & \equiv &
  \mathrm{dim}_q R
  \, ,
 \label{unknot_CS}
\end{eqnarray}
with
\begin{equation}
 [x]_q = q^{x/2} - q^{-x/2} \, .
\end{equation}
The denominator in (\ref{unknot_CS}) corresponds to
the partition function of $\U(N)$ Chern--Simons theory 
\begin{equation}
 \cZ_{\rm CS}(S^3;q)
  =
  \frac{e^{\frac{\pi i}{8}N(N-1)}}{N^{\frac{1}{2}}(k+N)^{\frac{N-1}{2}}}
  \prod_{i<j}^N
  \left(
   q^{\frac{1}{2}(-i+j)} - q^{-\frac{1}{2}(-i+j)}
  \right)
  \, .
\end{equation}

Let us now comment on the situation such that the representation of the
character obeys $d(\lambda) < N$.
In this case we do not obtain a simple determinantal formula, since the
Schur function with such a representation is not simply factorized any more.
For example, the character expectation value for the hook
representation, corresponding to the simplest situation $d(\lambda) =
1$, is given by
\begin{equation}
 \Big\langle
  W_{(\alpha|\beta)} (\unknot)
 \Big\rangle
 =
 \frac{1}{\cZ_{\rm ABJM}} \frac{1}{N!^2}
 \int [dx]^N [dy]^N \,
 \det \tilde{C}^2 \,
 \sum_{i,j=1}^N e^{\left(\alpha + \frac{1}{2} \right) x_i + \left(\beta
		    + \frac{1}{2}\right) y_j}
 \left(\tilde{C}^{-1}\right)_{ij}
 \, ,
 \label{WL_hook_rep}
\end{equation}
where we have
\begin{equation}
 \tilde{C} 
  =
  \left(
   \frac{1}{2 \cosh \frac{x_i-y_j}{2}}
  \right)_{1 \le i, j \le N}
  \, .
\end{equation}
Although it is difficult to find an explicit formula for this integral,
we expect that its asymptotic behavior is obtained from the determinantal
formula (\ref{unknot_NN01}) by taking the limit of $\alpha_1, \beta_1
\to \infty$.
Nevertheless, let us comment that we can apply the Fermi gas method even to this case.
Actually if we consider the grand canonical partition function, it turns
out to be written as a Fredholm determinant, due to the Giambelli
formula.
See, for example, \cite{Hatsuda:2013yua}.

\section{Torus knot matrix model}\label{sec:torus}

In addition to the unknot invariant, there is a similar integral formula
for the torus knot Wilson loop based on Chern--Simons
theory~\cite{Lawrence:1999CMP,Beasley:2009mb,Kallen:2011ny,Brini:2011wi},
\begin{equation}
 \Big\langle
  W_R(K_{P,Q})
 \Big\rangle
  =
  \frac{1}{\cZ_{\rm CS}^{(P,Q)}}
  \frac{1}{N!}
  \int \prod_{i=1}^N \frac{dx_i}{2\pi} \, e^{-\frac{1}{2\hat{g}_s} x_i^2}
  \prod_{i<j}^N
  \left(
   2 \sinh \frac{x_i-x_j}{2P}
  \right)
  \left(
   2 \sinh \frac{x_i-x_j}{2Q}
  \right)
  \mathrm{Tr}_R U(x)
  \, ,
  \label{CS_WL_torus}
\end{equation}
where the coupling constant is now rescaled $\hat{g}_s = PQ g_s$, and the
corresponding partition function is then given by
\begin{equation}
 \cZ_{\rm CS}^{(P,Q)}(S^3;q)
  =
  \frac{1}{N!}
  \int \prod_{i=1}^N \frac{dx_i}{2\pi} \, e^{-\frac{1}{2\hat{g}_s} x_i^2}
  \prod_{i<j}^N
  \left(
   2 \sinh \frac{x_i-x_j}{2P}
  \right)
  \left(
   2 \sinh \frac{x_i-x_j}{2Q}
  \right)
  \, .
  \label{CS_MM_torus}
\end{equation}
This is obtained from the ordinary matrix model by applying the
$\mathrm{SL}(2,\Z)$ transformation~\cite{Brini:2011wi}.
Note that it is also seen as the biorthogonal generalization of the
Chern--Simons matrix model~\cite{Dolivet:2006ii}.

For the torus knot invariant there is a useful formula, which is called
the Rosso--Jones formula~\cite{Rosso:1993JKTR} 
\begin{equation}
 \Big\langle
  W_R(K_{P,Q})
 \Big\rangle
 =
 \sum_V 
 c_{R,Q}^V \,
 \Big\langle
  W_V(K_{1,f})
 \Big\rangle
 \, ,
 \label{RJ_formula}
\end{equation}
with $f = P/Q$.
This means that the $(P,Q)$ torus knot invariant can be expressed as a
linear combination of the fractionally framed unknot invariant.
This formula is easily derived from the integral formula
(\ref{CS_WL_torus}) by using the Adams operation
\begin{equation}
 s_\lambda(u^Q)
  =
  \sum_{\mu} c_{\lambda,Q}^\mu \, s_\mu(u)
  \, .
  \label{Adams_op01}
\end{equation}
The coefficient $c_{\lambda,Q}^\mu$ can be determined by the Frobenius
formula for the Schur and power sum polynomials,
\begin{equation}
 s_\lambda
  =
  \sum_{\mu} \frac{1}{z_\mu} \chi_\lambda(C_\mu) \, p_\mu \, , \qquad
  p_\mu
  =
  \sum_\nu \chi_\nu (C_\mu) \, s_\nu  \, ,
  \label{Frobenius_formula}
\end{equation}
where $\chi_\lambda$ and $C_\mu$ are the character and the conjugacy
class for the symmetric group, and the coefficient $z_\mu$ is given by
$z_\mu = \prod_j \mu_j! \, j^{\mu_j}$.
Thus we have
\begin{equation}
 c_{\lambda,Q}^\mu 
  =
  \sum_{\nu} 
  \frac{1}{z_\nu} \chi_\lambda (C_\nu) \chi_\mu (C_{Q\nu})
  \, .
  \label{Adams_coeff}
\end{equation}
Since the Rosso--Jones formula (\ref{RJ_formula}) is obtained from the
representation theoretical point of view, it is natural to think that
there is a similar formula even for supergroup theories.
Actually supersymmetric polynomials obey the same Frobenius formula
(\ref{Frobenius_formula}) by definition~\cite{Bars:1982ps}:
Let $\U(x,y)$ be a diagonal $\U(N+M)$ matrix, and given an expansion of the $\U(N+M)$ character in terms of the power sum polynomial $\Tr U(x,y)^n$ as 
\begin{align}
 \Tr_R U(x,y)
 & =
 \sum_{R'} C_R^{R'} \, \prod_{i=1}^{\ell(R')} \Tr U(x,y)^{R'_i}
\end{align}
with $\ell(R)$ to be the number of non-zero entries in the Young diagram $R$, the supergroup character is just obtained by replacing $\Tr U(x,y)$ with $\Str U(x;y)$,
\begin{align}
 \Str_R U(x;y)
 & =
 \sum_{R'} C_R^{R'} \, \prod_{i=1}^{\ell(R')} \Str U(x;y)^{R'_i}
\end{align}
Therefore we obtain the Adams operation for $\U(N|M)$ theory
with the same coefficient (\ref{Adams_coeff}),
\begin{equation}
 s_\lambda(u^Q;v^Q)
  =
  \sum_{\mu} c_{\lambda,Q}^\mu \, s_\mu(u;v)
  \, .
  \label{Adams_op02}
\end{equation}

Following the above discussions, we now introduce a supermatrix
version of the torus knot matrix model (\ref{CS_MM_torus})
\begin{equation}
 \cZ_{\rm ABJM}^{(P,Q)}
  =
  \frac{1}{N!^2}
  \int [dx]^N [dy]^N \,
  \det \left( \frac{1}{2 \cosh \frac{x_i-y_j}{2P}}\right)
  \det \left( \frac{1}{2 \cosh \frac{x_i-y_j}{2Q}}\right)
  \, ,
  \label{ABJM_MM_torus}
\end{equation}
with the rescaled coupling constant $\hat{g}_s = PQ g_s$.%
\footnote{We can derive the so-called mirror description for this
partition function, as well as the ordinary ABJM matrix
model~\cite{Kapustin:2010xq}.
It depends on the parameters $(P,Q)$ in a trivial way
(\ref{ABJM_mirror_torus}).
See Appendix \ref{sec:mirror} for details.}
Then we consider the character expectation value with respect to
this partition function 
\begin{equation}
 \Big\langle
  W_R (K_{P,Q})
 \Big\rangle
   =
  \frac{1}{\cZ_{\rm ABJM}^{(P,Q)}}
  \frac{1}{N!^2}
  \int [dx]^N [dy]^N \,
  \det \left( \frac{1}{2 \cosh \frac{x_i-y_j}{2P}}\right)
  \det \left( \frac{1}{2 \cosh \frac{x_i-y_j}{2Q}}\right)
  \,
  s_\lambda (e^x;e^y)
  \, .
  \label{ABJM_WL_torus}
\end{equation}
Let us call this expectation value the torus knot character average.
We can easily show that this $(P,Q)$-deformed $\U(N|N)$
character average also satisfies the Rosso--Jones formula
(\ref{RJ_formula}) by applying the Adams operation for $\U(N|N)$ theory
(\ref{Adams_op02}).
In this sense it is enough to compute the framed unknot average to
obtain the torus knot average.
In fact, when we start with a generic partition
$\lambda=(\alpha_1,\cdots,\alpha_N|\beta_1,\cdots,\beta_N)$ for a
particular torus knot, representations appearing in the expansion
(\ref{RJ_formula}) only provide partitions satisfying $d(\lambda)=N$.%
\footnote{Although we do not have an explicit proof of this statement, we
check it with a number of examples by numerical calculations.}
Therefore we now focus on the torus knot and framed unknot with a
representation with $d(\lambda)=N$.
In this case we can show that the framed unknot average has the
same expression as (\ref{unknot_NN02}), up to the framing factor,
\begin{align}
 \Big\langle
  W_{R}(K_{1,f})
 \Big\rangle
 & =
 \frac{k^{-N}}{\cZ_{\rm ABJM}^{(1,1)}}
 q^{\frac{f}{2}\left( C_2(\mu) - C_2(\nu)\right)}
 \det_{1 \le i, j \le N}
 \left(
 \frac{1}{q^{\frac{1}{2}(\alpha_i+\beta_j+1)}
  + q^{-\frac{1}{2}(\alpha_i+\beta_j+1)}}
 \right)
 \nonumber \\
 & = 
 \frac{k^{-N}}{\cZ_{\rm ABJM}^{(1,1)}}
 q^{\frac{f}{2}\left( C_2(\mu) - C_2(\nu)\right)}
 \prod_{i,j=1}^N
 \left(
  q^{\frac{1}{2}(\alpha_i+\beta_j+1)}
  + q^{-\frac{1}{2}(\alpha_i+\beta_j+1)}
 \right)^{-1}
 \nonumber \\
 &  \hspace{4em}
  \times
  \prod_{i<j}^N
  \left(
   q^{\frac{1}{2}(\alpha_i-\alpha_j)} 
   - q^{-\frac{1}{2}(\alpha_i-\alpha_j)}
  \right)
  \left(
   q^{\frac{1}{2}(\beta_i-\beta_j)} 
   - q^{-\frac{1}{2}(\beta_i-\beta_j)}
  \right)
  \, .
  \label{framed_unknot_ABJM}
\end{align}
This means that the $\U(1|1)$ expectation value plays a role of the building
block for the torus knot average at least with this kind of representations.

For $\U(1|1)$ theory we can compute the character expectation value
(\ref{ABJM_WL_torus}) explicitly, as well as the unknot average
(\ref{WL_U(11)}). 
In this case we have
\begin{equation}
 \Big\langle
  W_{R}(K_{P,Q})
 \Big\rangle
 =
 \frac{4 \, q^{\frac{PQ}{2}(\alpha+\beta+1)(\alpha-\beta)}
       \left( q^{\frac{PQ}{2}(\alpha+\beta+1)} 
             + q^{-\frac{PQ}{2}(\alpha+\beta+1)}
       \right)}
      {\left( q^{\frac{P}{2}(\alpha+\beta+1)} 
	     + q^{-\frac{P}{2}(\alpha+\beta+1)}
       \right)
       \left( q^{\frac{Q}{2}(\alpha+\beta+1)} 
	     + q^{-\frac{Q}{2}(\alpha+\beta+1)}
       \right)}
 \, .       
\end{equation}
Then we obtain ``$\U(1|1)$ knot invariant'' for the
torus knot from this expectation value by implementing the normalization
with the unknot contribution (\ref{WL_U(11)}), as shown in
(\ref{WL_normalize}).
Removing the framing factor, we have
\begin{equation}
 J_R(K_{P,Q})
  =
 \frac{\left( q^{\frac{1}{2}(\alpha+\beta+1)} 
	     + q^{-\frac{1}{2}(\alpha+\beta+1)}
       \right)
       \left( q^{\frac{PQ}{2}(\alpha+\beta+1)} 
             + q^{-\frac{PQ}{2}(\alpha+\beta+1)}
       \right)}
      {\left( q^{\frac{P}{2}(\alpha+\beta+1)} 
	     + q^{-\frac{P}{2}(\alpha+\beta+1)}
       \right)
       \left( q^{\frac{Q}{2}(\alpha+\beta+1)} 
	     + q^{-\frac{Q}{2}(\alpha+\beta+1)}
       \right)}
 \, .
 \label{torus_U(11)}
\end{equation}
This is a generic formula for the $(P,Q)$ torus knot with the
representation $\lambda=(\alpha|\beta)$.
This expression is not a polynomial of $q$ and $q^{-1}$ in general, but
it is manifestly invariant under the exchange of $P
\leftrightarrow Q$, and also the inversion $q \leftrightarrow q^{-1}$.

\section{\texorpdfstring{$\U(N|M)$}{U(N|M)} theory}\label{sec:U(N|M)}

The argument shown above can be straightforwardly extended to
$\U(N|M)$ theory.
The $\U(N|M)$ supermatrix Chern--Simons model is obtained from the
Chern--Simons--matter theory with gauge group $\U(N)_k \times
\U(M)_{-k}$, which is called ABJ theory~\cite{Aharony:2008gk},
\begin{align}
 \cZ_{\rm ABJ}(S^3;q)
 & =
 \frac{1}{N!M!}
 \int [dx]^N [dy]^M 
 \prod_{i=1}^N \prod_{j=1}^M
 \left(
 2 \cosh \frac{x_i-y_j}{2}
 \right)^{-2}
 \nonumber \\
 &
 \hspace{5em} \times
 \prod_{i<j}^N
 \left(
 2 \sinh \frac{x_i-x_j}{2}
 \right)^2
 \prod_{i<j}^M
 \left(
 2 \sinh \frac{y_i-y_j}{2}
 \right)^2
 \, .
\end{align}
The matrix measure in this model can be also expressed as a determinant
using the generalized Cauchy determinant formula~\cite{Basor:1994MN}
\begin{equation}
 \Delta_N (u) \Delta_M (v)
  \prod_{i=1}^N \prod_{j=1}^M (u_i + v_j)^{-1}
  =
  \det \left(
	\begin{array}{c}
	 u_i^{k-1} \\ \left( u_i + v_j \right)^{-1}
	\end{array}
       \right)
  \, ,
\end{equation}
where we assume $N \ge M$, and the indices run as $i = 1, \cdots, N$,
$j=1, \cdots, M$ and $k=1, \cdots, N-M$.
Thus we have
\begin{align}
 &
 \prod_{i=1}^N \prod_{j=1}^M
 \left(
 2 \cosh \frac{x_i-y_j}{2}
 \right)^{-1}
 \prod_{i<j}^N 
 \left(
 2 \sinh \frac{x_i-x_j}{2}
 \right)
 \prod_{i<j}^M
 \left(
 2 \sinh \frac{y_i-y_j}{2}
 \right)
 \nonumber \\
 & =
 \prod_{i=1}^N e^{\frac{-N+M+1}{2} x_i}
 \prod_{j=1}^M e^{\frac{N-M+1}{2} y_j}
 \det
 \left(
 \begin{array}{c}
  e^{x_i (k-1)} \\ \left( e^{x_i} + e^{y_j} \right)^{-1}
 \end{array}
 \right)
 \, .
\end{align}

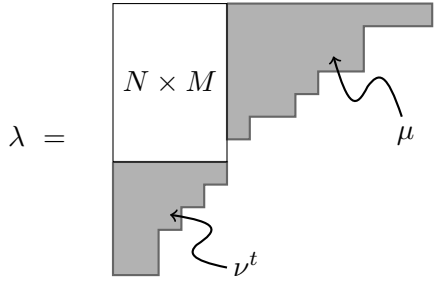
\begin{figure}[t]
 \begin{center}
  \begin{tikzpicture}
   \path (0,0.35) node {$\lambda \ =$};
   \path (1.75,1.05) node {$N \times M$};
   \path (4.85,0.35) node {$\mu$};
   \path (2.75,-1.35) node {$\nu^t$};

   \draw 
   (1,0) -- (1,2.1) -- (2.5,2.1) -- (2.5,0) -- (1,0) ;

   \filldraw [fill=gray,opacity=.6,draw=black,thick]
   (2.5,2.1) -- (5.2,2.1) -- (5.2,1.8) -- (4.3,1.8) -- 
   (4.3,1.2) -- (3.7,1.2) -- (3.7,0.9) -- (3.4,0.9) -- (3.4,0.6) --
   (2.8,0.6) -- (2.8,0.3) -- (2.5,0.3) -- (2.5,2.1);

   \filldraw [fill=gray,opacity=.6,draw=black,thick]
   (1,0) -- (1,-1.5) -- (1.6,-1.5) -- (1.6,-0.9) -- (1.9,-0.9)
   -- (1.9,-0.6) -- (2.2,-0.6) -- (2.2,-0.3) -- (2.5,-0.3) 
   -- (2.5,0) -- (1,0);

   \draw [<-,thick] 
   (3.9,1.4) .. controls (4.1,0.7) and (4.2,0.7) .. (4.35,1) 
   .. controls (4.5,1.2) and (4.6,1.2) .. (4.8,0.6) ;

   \draw [<-,thick]
   (1.75,-0.7) .. controls (2.4,-0.8) and (2.4,-0.9) .. (2.1,-1.1)
   .. controls (2,-1.2) and (2,-1.3) .. (2.5,-1.4);

  \end{tikzpicture}
 \end{center}
 \caption{The partition $\lambda = (14,11,11,9,8,6,5,5,4,3,2,2)$ 
 satisfying $\lambda_N \ge M$ with $N=7$ and $M=5$, which includes
 $\mu=(9,6,6,4,3,1,0)$ and $\nu^t=(5,4,3,2,2)$.
 }
 \label{fig:partition2}
\end{figure}

We then consider the expectation value of the Wilson loop operator with
respect to this partition function.
As in the case of $\U(N|N)$ theory, if the partition, corresponding to the
representation of the Wilson loop, satisfies $\lambda_N \ge M$ as
shown in Fig.~\ref{fig:partition2}, the Schur function is factorized
into the ordinary ones~\cite{Moens:2003JAC}
\begin{align}
 s_\lambda(u;v)
 & =
 s_\mu (u) \, s_\nu (v)
 \prod_{i=1}^N \prod_{j=1}^M
 (u_i + v_j)
 \nonumber \\
 & =
 \prod_{i=1}^N u_i^{\lambda_i+N-M-i}
 \prod_{j=1}^M v_j^{\lambda^t_j+M-N-j}
 \det
 \left(
 \begin{array}{c}
  u_i^{k-1} \\ \left( u_i + v_j \right)^{-1}
 \end{array}
 \right)
 \, ,
\end{align}
where the partitions $\mu$ and $\nu$ are defined as $\mu_i = \lambda_i -
M$ for $i = 1, \cdots, N$ and $\nu_j = \lambda^t_j - N$ for $j = 1,
\cdots, M$, or equivalently $\mu^t_i = \lambda^t_{i+M}$ and $\nu^t_j =
\lambda_{j+N}$.
Therefore the unknot character expectation value is now given by
\begin{align}
 \Big\langle
  W_R (\unknot)
 \Big\rangle
 & =
 \frac{1}{\cZ_{\rm ABJ}}
 \frac{1}{N! M!}
 \int [dx]^N [dy]^M
 \det
 \left(
  \begin{array}{c}
   e^{x_i (k-1)} \\ \left( e^{x_i} + e^{y_j} \right)^{-1}
  \end{array}
 \right)
 \det_N e^{x_i(\lambda_j-j+1)}
 \det_M e^{y_i(\lambda^t_j-j+1)}
 \nonumber \\
 & =
 \frac{1}{\cZ_{\rm ABJ}}
 \det
 \left(
 \begin{array}{c}
  \displaystyle
   \int \frac{dx}{2\pi} \, 
   e^{-\frac{1}{2g_s}x^2+x\left(\xi_i+k-\frac{1}{2}\right)} \\
  \displaystyle
   \int \frac{dx}{2\pi} \frac{dy}{2\pi} \,
   e^{-\frac{1}{2g_s}(x^2-y^2) + x \xi_i + y \eta_j}
   \left(
    2 \cosh \frac{x-y}{2}
   \right)^{-1}
 \end{array}
 \right)
 \, ,
\end{align}
with $\xi_i = \lambda_i - i + \frac{1}{2}$ for $i = 1, \cdots, N$ and
$\eta_j = \lambda^t_j - j + \frac{1}{2}$ for $j = 1, \cdots, M$.
This yields the determinantal formula for $\U(N|M)$ theory
\begin{align}
 \Big\langle
  W_R (\unknot)
 \Big\rangle
 & =
 \frac{1}{\cZ_{\rm ABJ}}
 \frac{i^{\frac{N-M}{2}}}{k^{\frac{N+M}{2}}}
 \prod_{i=1}^N q^{\frac{1}{2}(\xi_i^2 + \xi_i)}
 \prod_{j=1}^M q^{\frac{1}{2}(-\eta_j^2 + \eta_j)}
 \prod_{k=1}^{N-M} q^{\frac{1}{2}\left(k-\frac{1}{2}\right)^2}
 \det
 \left(
 \begin{array}{c}
  q^{\xi_i (k-1)} \\ \left( q^{\xi_i} + q^{\eta_j} \right)^{-1}
 \end{array}
 \right)
 \, .
\end{align}
This formula is also represented as follows,
\begin{align}
 &
 \Big\langle
  W_R (\unknot)
 \Big\rangle
 \nonumber \\
  & =
 \frac{1}{\cZ_{\rm ABJ}}
 \frac{i^{\frac{N-M}{2}}}{k^{\frac{N+M}{2}}} 
 q^{\frac{1}{6}(N-M)\left(N-M+\frac{1}{2}\right)\left(N-M-\frac{1}{2}\right)}
 \prod_{i=1}^N q^{\frac{1}{2} \left(\xi_i^2+(N-M)\xi_i\right)}
 \prod_{j=1}^M q^{-\frac{1}{2} \left(\eta_j^2+(N-M)\eta_j\right)}
 \nonumber \\
 &
 \times
 \prod_{i<j}^N 
 \left(
 q^{\frac{1}{2}\left(\xi_i-\xi_j\right)} 
 - q^{-\frac{1}{2}\left(\xi_i-\xi_j\right)}
 \right)
 \prod_{i<j}^M
 \left(
 q^{\frac{1}{2}\left(\eta_i-\eta_j\right)} 
 - q^{-\frac{1}{2}\left(\eta_i-\eta_j\right)}
 \right)
 \prod_{i=1}^N \prod_{j=1}^M
 \left(
 q^{\frac{1}{2}(\xi_i+\eta_j)} + q^{-\frac{1}{2}(\xi_i+\eta_j)}
 \right)^{-1}
 \, .
\end{align}
It is easy to see that this expression is reduced to the $\U(N|N)$
average (\ref{unknot_NN02}) and the $\U(N)$ invariant
(\ref{unknot_CS}) by taking $N=M$ and $M=0$, respectively.

We can similarly introduce the $\U(N|M)$ supermatrix model for torus knots
\begin{align}
 \cZ_{\rm ABJ}^{(P,Q)}(S^3;q)
 & =
 \frac{1}{N! M!} 
 \int [dx]^N [dy]^M 
 \prod_{i=1}^N \prod_{j=1}^M
 \left(
 2 \cosh \frac{x_i-y_j}{2P}
 \right)^{-1}
 \left(
 2 \cosh \frac{x_i-y_j}{2Q}
 \right)^{-1}
 \nonumber \\
 &
 \times
 \prod_{i<j}^N
 \left(
 2 \sinh \frac{x_i-x_j}{2P}
 \right)
 \left(
 2 \sinh \frac{x_i-x_j}{2Q}
 \right)
 \prod_{i<j}^M
 \left(
 2 \sinh \frac{y_i-y_j}{2P}
 \right)
 \left(
 2 \sinh \frac{y_i-y_j}{2Q}
 \right)
 \, .
 \label{ABJ_MM_torus}
\end{align}
The torus knot average is obtained from this matrix model by inserting the $\U(N|M)$ character, and thus satisfies the Rosso--Jones formula (\ref{RJ_formula}) thanks to the Adams operation (\ref{Adams_op02}).
We remark that in the limit $M \to 0$ this matrix model reproduces the integral formula for the ordinary $\U(N)$ torus knot invariant, which is the HOMFLY polynomial~\cite{Lawrence:1999CMP,Beasley:2009mb,Kallen:2011ny,Brini:2011wi}.
This supports a topological nature of the matrix model presented here.

In the next section we will discuss the spectral curve of the matrix model \eqref{ABJ_MM_torus} arising in the large $N$ limit.

\section{Spectral curve}\label{sec:sp_curve}

We then provide the spectral curve for the ABJ(M) matrix model for
$(P,Q)$ torus knots, which has been introduced in Sec.~\ref{sec:torus} and
Sec.~\ref{sec:U(N|M)}.
In order to obtain the spectral curve, we have to solve the saddle point
equations arising in the large $N$ limit of the corresponding matrix
model.
It is well known that the ABJM matrix model is perturbatively equivalent
to the Chern--Simons theory on the lens space $L(2,1) = S^3/\Z_2 \cong
\mathbb{RP}^3$, which is regarded as the two-cut solution of the
Chern--Simons matrix model~\cite{Marino:2009jd}.
As well as the ordinary ABJM matrix model, we can obtain the
spectral curve for the torus knot ABJM model from the
$(P,Q)$-modified Chern--Simons theory on the lens space $L(2,1)$
\begin{align}
 \cZ_{\rm CS}^{(P,Q)} (L(2,1);q)
  & = 
  \frac{1}{N!^2}
  \int \prod_{i=1}^N \frac{dx_i}{2\pi} \frac{dy_i}{2\pi} \,
  e^{-\frac{1}{2\hat{g}_s} (x_i^2 + y_i^2)} \,
  \prod_{i,j=1}^N
  \left(
   2 \cosh \frac{x_i-y_j}{2P}
  \right)
  \left(
   2 \cosh \frac{x_i-y_j}{2Q}
  \right)
  \nonumber \\
 &  \hspace{1em}
  \times
  \prod_{i<j}^N
  \left(
   2 \sinh \frac{x_i-x_j}{2P}
  \right)
  \left(
   2 \sinh \frac{x_i-x_j}{2Q}
  \right)
  \left(
   2 \sinh \frac{y_i-y_j}{2P}
  \right)
  \left(
   2 \sinh \frac{y_i-y_j}{2Q}
  \right)
 \, .
 \label{CS_MM_torus_lens}
\end{align}
This is interpreted as the two-cut matrix model of the original Chern--Simons torus knot model (\ref{CS_MM_torus}), whose gauge symmetry is broken as as $\U(N+N) \to \U(N) \times \U(N)$, or in general, $\U(N+M) \to \U(N) \times \U(M)$.


\begin{figure}[t]
 \begin{center}
  \begin{tikzpicture}

   \path (0,0) node {Chern--Simons matrix model (\ref{CS_MM})};
   \draw [thick,rounded corners] (-3.2,-0.5) rectangle (3.2,0.5);   

   \path (3.5,-3) node {ABJM matrix model (\ref{ABJM_MM})};
   \path (3.5,-3.5) node {(Chern--Simons supermatrix model)};
   \draw [thick,rounded corners] (0.3,-3.8) rectangle (6.7,-2.6);   

   \path (-5.1,-3) node {$(P,Q)$ torus knot};
   \path (-3.4,-3.5) node {matrix model (\ref{CS_MM_torus})};
   \draw [thick,rounded corners] (-7,-2.6) rectangle (-1.5,-3.8);

   \path (0,-6.5) node {Torus knot ABJM model (\ref{ABJM_MM_torus})};
   \draw [thick,rounded corners] (-3.2,-7) rectangle (3.2,-6);

   \draw [->,ultra thick,magenta] (2,-0.6) -- (3.5,-2.5);
   \draw [->,ultra thick,cyan] (-2,-0.6) -- (-3.5,-2.5);
   \draw [->,ultra thick,cyan] (3.5,-3.9) -- (2,-5.8);
   \draw [->,ultra thick,magenta] (-3.5,-3.9) -- (-2,-5.8);

   \path (3,-1.5) node [right] {2-cut \& analytic continuation};
   \path (-3,-1.5) node [left] {$\mathrm{SL}(2,\Z)$ transform};

   \path (3,-5) node [right] {$\mathrm{SL}(2,\Z)$ transform};
   \path (-3,-5) node [left] {2-cut \& analytic continuation};

  \end{tikzpicture}
 \end{center}
 \caption{Two ways of obtaining the $(P,Q)$ torus knot ABJM
 (supermatrix) model.}
 \label{flowchart}
\end{figure}
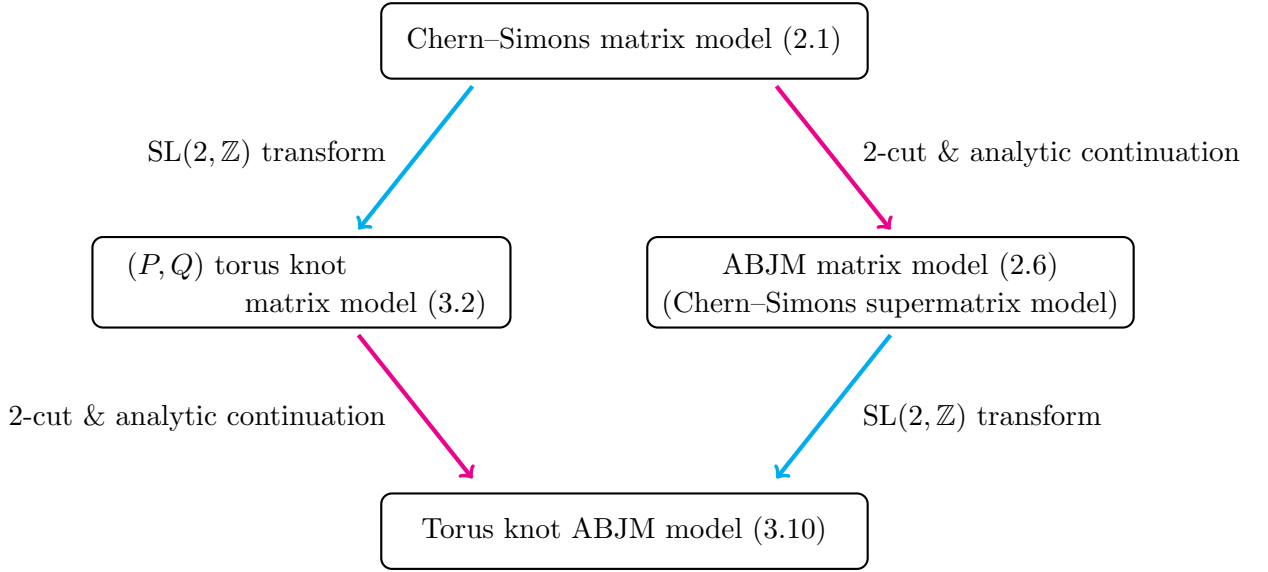

As shown in Fig.~\ref{flowchart}, we have two ways of obtaining the
spectral curve for the torus knot ABJM matrix model.
The first is to apply the $\SL(2,\Z)$ transformation to the spectral curve
for ABJM theory, which is given by that for the lens space
Chern--Simons theory, through the analytic
continuation~\cite{Aganagic:2002wv,Halmagyi:2003ze}.
We note that this kind of symplectic transformation is considered to
discuss the knot invariant for the lens
space~\cite{Jockers:2012pz,Stevan:2013tha}.
The second is directly solving the saddle point equation for the matrix model
(\ref{CS_MM_torus_lens}).
In this case its saddle point analysis is similar to the ordinary
one-cut solution, which is discussed in \cite{Brini:2011wi}.
We will show that these two methods provide a consistent result.

\subsection{Symplectic transformation}\label{sec:symplectic}

Let us start with the spectral curve for the Chern--Simons theory on the
lens space $S^3/\Z_2 \cong \mathbb{RP}^3$, which is essentially
equivalent to that for ABJM theory.
The spectral curve $\cC$ is defined as the zero locus of the two-parameter
function
\begin{equation}
 \cC = \Big\{ (U,V) \in 
  \C^* \times \C^* \, \Big| \, H(U,V) = 0 \Big\} \, ,
\end{equation}
where the function $H(U,V)$ for the lens space $L(2,1)$ is now given
by~\cite{Halmagyi:2003ze}
\begin{equation}
 H(U,V) 
  = 
  c \left(
     U + \frac{V^2}{U}
    \right)
  - V^2 + \zeta V - 1
  \, .
  \label{ABJM_curve00}
\end{equation}
The parameter $\zeta$ is to be determined,
and $c$ is related to the 't Hooft coupling constant as $c =
\exp g_s (N+M)/2$ for the two-cut model with $\U(N) \times \U(M)$ symmetry.%
\footnote{%
The spectral curve for Chern--Simons theory on the lens space $L(r,1) = S^3/\Z_r$ (the $r$-cut Chern--Simons matrix model) is given by~\cite{Halmagyi:2003ze}
\begin{equation}
 H
  (U,V)
  =
  c \left( U + \frac{V^r}{U} \right)
  - p_r(V) = 0
  \, ,
  \label{Lens_curve01}
\end{equation}
where $p_r(V)$ is a degree $r$ polynomial such that the coefficients of $V^r$ and $V^0$ are given by one, $p_r(V) = V^r + \cdots + 1$.
When the Chern--Simons gauge group is broken as $\U(N_1 + \cdots + N_r) \to \U(N_1) \times \cdots \times \U(N_r)$, the parameter $c$ corresponds to the total 't Hooft coupling $c = \exp g_s (N_1 + \cdots + N_r)/2$.
Since the polynomial $p_r(V)$ has $r-1$ parameters, the total number of the parameters becomes $1 + (r-1) = r$, which is consistent with that of the subgroups, $\U(N_i)$ with $i = 1, \cdots, r$.}

Although the algebraic curve characterized by the function \eqref{ABJM_curve00} is naturally obtained from the saddle point analysis of the matrix model, we replace the variable $(U,V) \to (U^2, c^{-\frac{1}{2}} U^{\frac{1}{2}} V)$ to obtain a homogeneous expression in $U$ and $V$,
\begin{equation}
 H(U,V) 
  =
  c U^2 + V^2 - c^{-1} V^2 U^2 + c^{-\frac{1}{2}} \zeta V U - 1 = 0
  \, .
\end{equation}
This solves
\begin{equation}
 V =
  \frac{1}{1-c^{-1}U^2}
  \left(
   -\frac{\zeta}{2 c^{\frac{1}{2}}} U
   \pm
   \sqrt{ \frac{\zeta^2}{4c} U^2 +
          (1 - c U^2)(1 - c^{-1} U^2)}
  \right)
  \, ,
  \label{ABJM_curve02}
\end{equation}
which is a more convenient expression to perform the contour integral as discussed later.
By taking the limit $\zeta \to 0$, this solution is just reduced to the
one-cut solution, which reproduces the previous result~\cite{Brini:2011wi}, up to a proper replacement of the variable $V \to V^2$,
\begin{equation}
 V \ \stackrel{\zeta \to 0}{\longrightarrow} \
  \frac{1 - c U^2}{1 - c^{-1} U^2}
  \, .
\end{equation}
If we write the discriminant of (\ref{ABJM_curve02}) as
\begin{equation}
 \frac{\zeta^2}{4c} U^2 + (1 - c U^2)(1 - c^{-1} U^2)
  =
  \Big( U^2 - \alpha \Big) \Big( U^2 - \frac{1}{\alpha} \Big)
  \, ,
\end{equation}
the end point of the cut is determined by
\begin{equation}
 \alpha + \frac{1}{\alpha} = c + \frac{1}{c} - \frac{\zeta^2}{4c}
  \, .
\end{equation}
Thus the parameter $\zeta$ is seen as the blow-up parameter for the
spectral curve, which is determined by requiring that the filling
fractions of the cuts are:
\begin{align}
\frac{1}{2\pi i} \oint_{[\alpha^{-1/2},\alpha^{1/2}]} \log{V} \, \frac{dU}{U}  
 & = g_s N  \, , \label{FF01} \\
\frac{1}{2\pi i} \oint_{[-\alpha^{1/2},-\alpha^{-1/2}]} \log{V} \, \frac{dU}{U}
 & = g_s M  \, , \label{FF02}
\end{align}
for $\U(N) \times \U(M)$ theory where the integration contour surrounds the corresponding segments counterclockwise.

We can obtain the spectral curve for the $(P,Q)$ torus knot from
(\ref{ABJM_curve02}) through the symplectic transformation, which is
characterized by the $\SL(2,\Z)$ matrix~\cite{Brini:2011wi}
\begin{equation}
 M_{P,Q}
  =
  \left(
   \begin{array}{cc}
    Q & P \\ \gamma & \delta
   \end{array}
  \right)
  \, ,
  \label{SL_matrix}
\end{equation}
where these integer entries satisfy the condition
\begin{equation}
 Q \delta - P \gamma = 1 \, .
\end{equation}
We then apply the following choice of variables corresponding to the
matrix (\ref{SL_matrix}),
\begin{align}
 X & = U^Q V^P \, , \\
 Y & = U^\gamma V^\delta \, .
\end{align}
In this case by substituting the original expression (\ref{ABJM_curve02})
and rescaling the variable $U^2 \to c^{\frac{P}{Q}} U^2$, we obtain the
spectral curve for the $(P,Q)$ torus knot Chern--Simons theory on the
lens space
\begin{align}
 X & = 
  \frac{U^{Q}}{\left(1-c^{\frac{P}{Q}-1} U^2\right)^P}
  \left(
   - \frac{\zeta}{2} c^{(\frac{P}{Q}-1)/2} U
   \pm 
   \sqrt{
   \frac{\zeta^2}{4} c^{\frac{P}{Q}-1} U^2
   + \left(
      1 - c^{\frac{P}{Q}-1} U^2
     \right)
     \left(
      1 - c^{\frac{P}{Q}+1} U^2
     \right)
        }
  \right)^P
 \, , \label{ABJM_torus_curve01} \\
 V & =
  \frac{1}{1-c^{\frac{P}{Q}-1} U^2}
  \left(
   - \frac{\zeta}{2} c^{(\frac{P}{Q}-1)/2} U
   \pm 
   \sqrt{
   \frac{\zeta^2}{4} c^{\frac{P}{Q}-1} U^2
   + \left(
      1 - c^{\frac{P}{Q}-1} U^2
     \right)
     \left(
      1 - c^{\frac{P}{Q}+1} U^2
     \right)
        }
  \right)
 \, . \label{ABJM_torus_curve02}
\end{align}
It is shown in~\cite{Jockers:2012pz,Stevan:2013tha} that the new curve
obtained through this kind of symplectic transformation gives the
topological invariants for torus knots in the lens space $S^3/\Z_2$, which is dual
to the topological string on the local $\mathbb{P}^1 \times
\mathbb{P}^1$ geometry.

\subsection{Saddle point analysis}

We study the torus knot spectral curve from the large $N$ limit
of the matrix model (\ref{CS_MM_torus_lens}), 
and then check its
consistency with the result obtained through the symplectic
transformation, (\ref{ABJM_torus_curve01}) and (\ref{ABJM_torus_curve02}).
We now rewrite the matrix integral (\ref{CS_MM_torus_lens}) with another
set of variables, $u_i = e^{x_i/(PQ)}$ and $v_i = e^{y_i/(PQ)}$,
\begin{align}
 \cZ_{\rm CS}^{(P,Q)} (L(2,1);q) 
 & =
 \frac{1}{N!^2} \int \! \frac{d^N u}{(2\pi)^N} \frac{d^N v}{(2\pi)^N} \,
 \exp \Bigg[ \sum_{i,j=1}^N 
 \left(
 \log (u_i^P + v_j^P) + \log (u_i^Q + v_j^Q)
 \right)
 \nonumber \\
 & \hspace{1em}
 + \sum_{i<j}^N
 \left(
 \log (u_i^P - u_j^P) + \log (u_i^Q - u_j^Q)
 + \log (v_i^P - v_j^P) + \log (v_i^Q - v_j^Q)
 \right)
 \nonumber \\
 & \hspace{1em}
 - \sum_{i=1}^N
 \left(
 \frac{PQ}{2g_s} \left( \log^2 u_i + \log^2 v_i \right)
 + \left( \frac{P+Q}{2} (2N-1) + 1 \right) \left( \log u_i + \log v_i \right)
 \right)
 \Bigg]
 \, .
\end{align}
The matrix integral has a convex potential (see \cite{Borot:2013lpa}), and this implies that the integrand has a unique minimum.

In this case we have two saddle point equations with $u_i$ and $v_i$ variables,
\begin{align}
 \sum_{j(\neq i)}^N
 \left[
 \frac{P u_i^P}{u_i^P - u_j^P}
 + \frac{Q u_i^Q}{u_i^Q - u_j^Q}
 \right]
 +
 \sum_{j=1}^N
 \left[
 \frac{P u_i^P}{u_i^P + v_j^P}
 + \frac{Q u_i^Q}{u_i^Q + v_j^Q}
 \right]
 & =
 \frac{PQ}{g_s} \log u_i
 + \frac{P+Q}{2} (2N-1) + 1
 \, , \\
 \sum_{j(\neq i)}^N
 \left[
 \frac{P v_i^P}{v_i^P - v_j^P}
 + \frac{Q v_i^Q}{v_i^Q - v_j^Q}
 \right]
 +
 \sum_{j=1}^N
 \left[
 \frac{P v_i^P}{v_i^P + u_j^P}
 + \frac{Q v_i^Q}{v_i^Q + u_j^Q}
 \right]
 & =
 \frac{PQ}{g_s} \log v_i
 + \frac{P+Q}{2} (2N-1) + 1
 \, .
\end{align}
Let us now rewrite these equations in terms of the
resolvent~\cite{Brini:2011wi}.
Using the formula
\begin{align}
 \frac{P x^{P-1}}{x^P - y^P}
  & =
  \sum_{k=0}^{P-1}
  \frac{1}{x - \omega^{-kQ} y}
  \, , \\
 \frac{P x^{P-1}}{x^P + y^P}
  & =
  \sum_{k=0}^{P-1}
  \frac{1}{x - \omega^{-(k+\frac{1}{2})Q} y} 
  \, ,
\end{align}
with the primitive $PQ$-th root of unity $\omega = \exp 2 \pi i/(PQ)$, we have
\begin{align}
 &
 \sum_{j(\neq i)}^N \frac{P u_i^{P}}{u_i^P - u_j^P}
 + \sum_{j=1}^N \frac{P u_i^P}{u_i^P + v_j^P}
 \nonumber \\
 & =
 \sum_{j(\neq i)}^N \frac{u_i}{u_i - u_j}
 + \sum_{k=1}^{P-1}
 \left(
 \frac{1}{g_s} W_0^{(1)}(u_i \, \omega^{kQ}) - \frac{1}{1-\omega^{-kQ}}
 \right)
 + \sum_{k=0}^{P-1}
 \frac{1}{g_s} W_0^{(2)}(u_i \, \omega^{(k+\frac12)Q)})
 \, ,
\end{align}
where $W_0^{(i)}(u)$ is the leading contribution to the resolvents
\begin{align}
 W^{(1)}(u) 
 & =
 \left\langle
 g_s \sum_{i=1}^N \frac{u}{u-u_i}
 \right\rangle 
 = \sum_{g=0}^\infty g_s^{2g-1} W_g^{(1)} (u)
 \, , \\
 W^{(2)}(u) 
 & =
 \left\langle
 g_s \sum_{i=1}^N \frac{u}{u-v_i}
 \right\rangle 
 = \sum_{g=0}^\infty g_s^{2g-1} W_g^{(2)} (u)
 \, .
\end{align}
These resolvents are analytic in the complex plane except for a finite
set of cuts.
We now assume that each resolvent has only one cut in the complex plane $\C^*$. We can check later that this assumption is correct, using unicity of the extremum guaranteed by \cite{Borot:2013lpa}.

We introduce the 't Hooft coupling constants for this matrix model.
If we consider a generic situation with the gauge group $\U(N_1+N_2) \to
\U(N_1) \times \U(N_2)$, we have two constants
\begin{equation}
 t^{(i)} = g_s N_i
  \, , \qquad
  i = 1, \, 2 \, .
\end{equation}
The summation of them is denoted by $t=t^{(1)}+t^{(2)}$.
ABJM theory corresponds to the situation such that $N_2$ is
analytically continuated as $N_2 \to -N_2$, and then the gauge group
ranks are chosen to be $N_{1,2} \to N$.
This means that the total 't Hooft coupling becomes zero $t = 0$.
Then the boundary conditions for the resolvents are given by
\begin{equation}
 W_0^{(i)}(u)
  \ \longrightarrow \
  \begin{cases}
   0 & (u \to 0) \\
   t^{(i)} & (u \to \infty) 
  \end{cases}
  \, ,
\end{equation}
and the saddle point equations in the 't Hooft limit with $N \to \infty$
and $g_s \to 0$ yield
\begin{align}
 PQ \log u + \frac{P+Q}{2} t
 & =
 W_0^{(1)}(u+i0) + W_0^{(1)}(u-i0)
 + \sum_{k=1}^{P-1} W_0^{(1)}(u\,\omega^{kQ})
 + \sum_{k=1}^{Q-1} W_0^{(1)}(u\,\omega^{kP})
 \nonumber \\
 & \hspace{5em}
 + \sum_{k=0}^{P-1} W_0^{(2)}(u\,\omega^{(k+\frac12)Q})
 + \sum_{k=0}^{Q-1} W_0^{(2)}(u\,\omega^{(k+\frac12)P})
 \, ,
 \label{saddle_pt01}
\end{align}
\begin{align}
 PQ \log u + \frac{P+Q}{2} t
 & =
 W_0^{(2)}(u+i0) + W_0^{(2)}(u-i0)
 + \sum_{k=1}^{P-1} W_0^{(2)}(u\,\omega^{kQ})
 + \sum_{k=1}^{Q-1} W_0^{(2)}(u\,\omega^{kP})
 \nonumber \\
 & \hspace{5em}
 + \sum_{k=0}^{P-1} W_0^{(1)}(u\,\omega^{(k+\frac12)Q})
 + \sum_{k=0}^{Q-1} W_0^{(1)}(u\,\omega^{(k+\frac12)P})
 \, .
 \label{saddle_pt02}
\end{align}
In order to deal with these equations, it is convenient to
introduce the exponentiated resolvents
\begin{align}
 y^{(a)}(u)
 & =
 - u \exp \frac{P+Q}{PQ}
 \left(
 \frac{t}{2} - W_0^{(1)} (u) - W_0^{(2)} (u \, \omega^{\frac12 Q})
 \right)
 \, , \\
 y^{(b)}(u)
 & =
 - u \exp \frac{P+Q}{PQ}
 \left(
 \frac{t}{2} - W_0^{(1)} (u) - W_0^{(2)} (u \, \omega^{\frac12 P})
 \right)
 \, ,
\end{align}
with the boundary behavior
\begin{equation}
 y^{(a,b)}(u)
  \ \longrightarrow \
  \begin{cases}
   - u \, e^{\frac{P+Q}{2PQ} t} & (u \to 0) \\
   - u \, e^{-\frac{P+Q}{2PQ} t} & (u \to \infty) \\
  \end{cases}
  \, .
\end{equation}
The saddle point equations are now written as follows,
\begin{align}
 y^{(a)}(u+i0) y^{(b)}(u-i0)
 \prod_{k=1}^{P-1} y^{(a)} (u \, \omega^{kQ})
 \prod_{k=1}^{Q-1} y^{(b)} (u \, \omega^{kP})
 & = 1 \, , 
 \label{saddle_pt03} \\
 y^{(a)} ((u+i0) \omega^{\frac{1}{2}Q}) 
 y^{(b)} ((u-i0) \omega^{\frac{1}{2}P})
 \prod_{k=1}^{P-1} y^{(a)} (u \, \omega^{(k+\frac{1}{2})Q})
 \prod_{k=1}^{Q-1} y^{(b)} (u \, \omega^{(k+\frac{1}{2})P})
 & = 1 \, .
 \label{saddle_pt04}
\end{align}
Because it is converted to each other under the exchange of $W^{(1)}(u)$ and
$W^{(2)}(u)$, these equations imply an equivalent condition for the
resolvents, which is essentially the same as the one-cut Chern--Simons matrix
model for the $(P,Q)$ torus knot~\cite{Brini:2011wi}.
Thus we can apply a similar approach to solve these equations.

\begin{table}[t]
 \begin{center}
 \begin{tikzpicture}[scale=0.9]

  \foreach \x in {0,1,2}
  {
  \foreach \y in {0,1}
  {

  \draw[opacity=0] (3.5*\x,0) 
  -- +(\y*360/2+\x*360/3:0.4) coordinate (i1)
  -- +(\y*360/2+\x*360/3:1) coordinate (e1)
  -- +(\y*360/2+\x*360/3:1.4) coordinate (c1\x\y);

  \draw[opacity=0] (3.5*\x,0) 
  -- +(\y*360/2+\x*360/3+360/4:0.4) coordinate (i2)
  -- +(\y*360/2+\x*360/3+360/4:1) coordinate (e2)
  -- +(\y*360/2+\x*360/3+360/4:1.4) coordinate (c2\x\y);

  \draw[thick] (i1) -- (e1);
  \draw[thick,densely dotted] (i2) -- (e2) ;

  \filldraw (i1) circle [radius=1pt];
  \filldraw (e1) circle [radius=1pt];
  \filldraw (i2) circle [radius=1pt];
  \filldraw (e2) circle [radius=1pt];

  }

  \draw (c1\x0) circle [radius=0] node {$F_3$};
  \draw (c1\x1) circle [radius=0] node {$F_4$};
  \draw (c2\x0) circle [radius=0] node {$F_3'$};
  \draw (c2\x1) circle [radius=0] node {$F_4'$};

  \draw (3.5*\x,2.5) circle [radius=0] 
  node {$F_\x$};

  }

  \foreach \x in {0,1}
  {
  \foreach \y in {0,1,2}
  {

  \draw[opacity=0] (3.5*\x+10.5,0) 
  -- +(\y*360/3+\x*360/2:0.4) coordinate (i1)
  -- +(\y*360/3+\x*360/2:1) coordinate (e1)
  -- +(\y*360/3+\x*360/2:1.4) coordinate (c1\x\y);

  \draw[opacity=0] (3.5*\x+10.5,0) 
  -- +(\y*360/3+\x*360/2+360/6:0.4) coordinate (i2)
  -- +(\y*360/3+\x*360/2+360/6:1) coordinate (e2)
  -- +(\y*360/3+\x*360/2+360/6:1.4) coordinate (c2\x\y);

  \draw[thick] (i1) -- (e1);
  \draw[thick,densely dotted] (i2) circle -- (e2) ;

  \filldraw (i1) circle [radius=1pt];
  \filldraw (e1) circle [radius=1pt];
  \filldraw (i2) circle [radius=1pt];
  \filldraw (e2) circle [radius=1pt];

  }

  \draw (c1\x0) circle [radius=0] node {$F_0$};
  \draw (c1\x1) circle [radius=0] node {$F_1$};
  \draw (c1\x2) circle [radius=0] node {$F_2$};
  \draw (c2\x0) circle [radius=0] node {$F_0'$};
  \draw (c2\x1) circle [radius=0] node {$F_1'$};
  \draw (c2\x2) circle [radius=0] node {$F_2'$};

  }

  \draw (10.5,2.5) circle [radius=0] 
  node {$F_3$};
  \draw (14,2.5) circle [radius=0] 
  node {$F_4$};

  \draw (-1.5,2) -- (15.5,2);
  \draw[double distance=1.5pt] (-1.5,3) -- (15.5,3);
  \draw[double distance=1.5pt] (-1.5,-2) -- (15.5,-2);

 \end{tikzpicture}
 \end{center}
 \begin{center}
 \begin{tikzpicture}[scale=0.9]

  \foreach \x in {0,1,2}
  {
  \foreach \y in {0,1}
  {

  \draw[opacity=0] (3.5*\x,0) 
  -- +(\y*360/2+\x*360/3-360/4+360/6:0.4) coordinate (i1)
  -- +(\y*360/2+\x*360/3-360/4+360/6:1) coordinate (e1)
  -- +(\y*360/2+\x*360/3-360/4+360/6:1.4) coordinate (c1\x\y);

  \draw[opacity=0] (3.5*\x,0) 
  -- +(\y*360/2+\x*360/3+360/6:0.4) coordinate (i2)
  -- +(\y*360/2+\x*360/3+360/6:1) coordinate (e2)
  -- +(\y*360/2+\x*360/3+360/6:1.4) coordinate (c2\x\y);

  \draw[thick] (i1) -- (e1);
  \draw[thick,densely dotted] (i2) -- (e2) ;

  \filldraw (i1) circle [radius=1pt];
  \filldraw (e1) circle [radius=1pt];
  \filldraw (i2) circle [radius=1pt];
  \filldraw (e2) circle [radius=1pt];

  }

  \draw (3.5*\x,2.5) circle [radius=0] 
  node {$F'_\x$};  

  \draw (c1\x0) circle [radius=0] node {$F_4'$};
  \draw (c1\x1) circle [radius=0] node {$F_3'$};
  \draw (c2\x0) circle [radius=0] node {$F_3$};
  \draw (c2\x1) circle [radius=0] node {$F_4$};

  }

  \foreach \x in {0,1}
  {
  \foreach \y in {0,1,2}
  {

  \draw[opacity=0] (3.5*\x+10.5,0) 
  -- +(\y*360/3+\x*360/2-360/6+360/4:0.4) coordinate (i1)
  -- +(\y*360/3+\x*360/2-360/6+360/4:1) coordinate (e1)
  -- +(\y*360/3+\x*360/2-360/6+360/4:1.4) coordinate (c1\x\y);

  \draw[opacity=0] (3.5*\x+10.5,0) 
  -- +(\y*360/3+\x*360/2+360/4:0.4) coordinate (i2)
  -- +(\y*360/3+\x*360/2+360/4:1) coordinate (e2)
  -- +(\y*360/3+\x*360/2+360/4:1.4) coordinate (c2\x\y);

  \draw[thick] (i1) -- (e1);
  \draw[thick,densely dotted] (i2) circle -- (e2) ;

  \filldraw (i1) circle [radius=1pt];
  \filldraw (e1) circle [radius=1pt];
  \filldraw (i2) circle [radius=1pt];
  \filldraw (e2) circle [radius=1pt];

  }

  \draw (c1\x0) circle [radius=0] node {$F_2'$};
  \draw (c1\x1) circle [radius=0] node {$F_0'$};
  \draw (c1\x2) circle [radius=0] node {$F_1'$};
  \draw (c2\x0) circle [radius=0] node {$F_0$};
  \draw (c2\x1) circle [radius=0] node {$F_1$};
  \draw (c2\x2) circle [radius=0] node {$F_2$};

  }

  \draw (10.5,2.5) circle [radius=0] 
  node {$F_3'$};
  \draw (14,2.5) circle [radius=0] 
  node {$F_4'$};

  \draw (-1.5,2) -- (15.5,2);
  \draw[double distance=1.5pt] (-1.5,3) -- (15.5,3);
  \draw[double distance=1.5pt] (-1.5,-2) -- (15.5,-2);


 \end{tikzpicture}
 \end{center}

 \caption{The cuts of the functions $F_k(u)$ and $F_{Q+l}(u)$ for
 $(P,Q)=(2,3)$, where $F_k = F_k(u)$,
 $F_k'=F_k(u\,\omega^{\frac{1}{2}(P-Q)})$ and
 $F_{Q+l}'=F_{Q+l}(u\,\omega^{\frac{1}{2}(Q-P)})$.
 The solid and dotted lines correspond to the cuts from the first
 resolvent $W^{(1)}(u)$ and the second resolvent $W^{(2)}(u)$.
 For example, we can see $F_0 = F_3$, $F_4$ under crossing the
 corresponding cut of $W^{(1)}(u)$, and $F_0 = F_3'$, $F_4'$ through
 the cut from $W^{(2)}(u)$.}
 \label{cuts}
\end{table}

We consider the products of the resolvents
\begin{align}
 F_k (u)
 & =
 \prod_{l=0}^{P-1} y^{(a)} (u \, \omega^{kP+lQ})
 \, , \quad 0 \le k \le Q-1 \, ,\\
 F_{Q+l} (u)
 & =
 \prod_{k=0}^{Q-1} \frac{1}{y^{(b)} (u \, \omega^{kP+lQ})}
 \, , \quad 0 \le l \le P-1 \, .
\end{align}
Since we have assumed that the original resolvents $W^{(i)}(u)$ have a
single cut, these functions $F_k (u)$ and $F_{Q+l}(u)$ have $2P$ and $2Q$ cuts,
obtained by rotating the original one with integer multiple angles of
$2\pi/(PQ)$.
The total number of the cuts is thus $2PQ$.
Due to the saddle point equations
they satisfy
\begin{align}
 \begin{array}{rclc}
 F_k (u-i0) & = & F_{Q+l} (u+i0) & \mbox{for} \quad W^{(1)}(u) \, , \\
 F_k (u-i0) & = & F_{Q+l} ((u+i0)\omega^{\frac{1}{2}(Q-P)}) &
 \mbox{for} \quad W^{(2)}(u) \, ,\\
 F_k ((u-i0)\omega^{\frac{1}{2}(P-Q)}) 
 & = & F_{Q+l} ((u+i0)\omega^{\frac{1}{2}(Q-P)}) &
 \mbox{for} \quad W^{(1)}(u) \, , \\
 F_k ((u-i0)\omega^{\frac{1}{2}(P-Q)}) & = & F_{Q+l} (u+i0) &
 \mbox{for} \quad W^{(2)}(u) \, .
 \end{array}
\end{align}
This means that
$F_k (u-i0) = F_{Q+l} (u+i0)$ under crossing the cut from the first
resolvent $W^{(1)}(u)$, $F_k (u-i0) = F_{Q+l} ((u+i0)
\omega^{\frac{1}{2}(Q-P)})$ for the cut from the second $W^{(2)}(u)$,
and so on.
See Table~\ref{cuts} for the case with $(P,Q)=(2,3)$.

Using these functions we define a function
\begin{equation}
 \mathcal{S} (u,f)
  = \mathcal{S}_1 (u,f) \, \mathcal{S}_2 (u,f) \, ,
\end{equation}
where
\begin{align}
 \mathcal{S}_1(u,f)
  & =
  \prod_{k=0}^{Q-1}
  \Big( f - F_k (u) \Big)
  \prod_{l=0}^{P-1}
  \Big( f - F_{Q+l} (u) \Big)
  \, , \\
 \mathcal{S}_2(u,f)
  & =
  \prod_{k=0}^{Q-1}
  \left( f - F_k (u \, \omega^{\frac{1}{2}(P-Q)}) \right)
  \prod_{l=0}^{P-1}
  \left( f - F_{Q+l} (u \, \omega^{\frac{1}{2}(Q-P)}) \right)
  \, .
\end{align}
This function has no cut in the complex plane
\begin{equation}
 \mathcal{S} (u-i0, f) =  \mathcal{S} (u+i0, f) \, ,
\end{equation}
and the only singularities at $u = 0$ or $u = \infty$ as poles.
This implies that $S(u,f)$ is an entire function of $u$ in $\mathbb C^*$ with polynomial behavior at $0$ and at $\infty$, therefore it must be a Laurent polynomial of $u$.
Moreover since this function satisfies $\mathcal{S}(u,f) = \mathcal{S}(u \,
\omega, f)$, it depends only on $u^{PQ}$, and must be a Laurent polynomial of $u^{PQ}$.
We remark that $\mathcal{S}_1(u,f)$ and $\mathcal{S}_2(u,f)$ still have
the cuts, and thus they are not analytic functions.

By definition, ${\cal S}(u,f)$ vanishes when $f=F_k(u)$, and thus the spectral curve is the algebraic equation:
\begin{equation}
 \mathcal{S}(u,f)  = 0 \, .
\end{equation}
As shown in \cite{Brini:2011wi} we can determine coefficients of the polynomial
$\mathcal{S}(u,f)$ by the asymptotic behavior of $F_k(u)$ and
$F_{Q+l}(u)$, 
\begin{align}
 F_k(u)
  & \, \longrightarrow \,
  \begin{cases}
   - \omega^{kP^2} \, e^{\frac{(P+Q)}{2Q} t} \, u^{P} & ( u \to 0 ) \\
   - \omega^{kP^2} \, e^{-\frac{(P+Q)}{2Q} t} \, u^{P} & ( u \to \infty )
  \end{cases}
 \, , 
 \label{F_bdry01}\\
 F_{Q+l}(u)
  & \, \longrightarrow \,
  \begin{cases}
   - \omega^{-lQ^2} \, e^{-\frac{(P+Q)}{2P} t} \, u^{-Q} & ( u \to 0 ) \\
   - \omega^{-lQ^2} \, e^{\frac{(P+Q)}{2P} t} \, u^{-Q} & ( u \to \infty )
  \end{cases}
 \, .
 \label{F_bdry02} 
\end{align}
If we write
\begin{equation}
 \mathcal{S}_{i} (u,f) 
  =
  \sum_{k=0}^{P+Q} (-1)^k \, \mathcal{S}_{i,k}(u) \, f^{P+Q-k} 
  \, , \quad i = 1, \, 2 \, ,
\end{equation}
this behavior implies
\begin{align}
 &
 \begin{cases}
  \mathcal{S}_{i,k}(u) = \cO (u^{-kQ}) \,,
  & 1 \le k \le P-1 \, , \\
  \mathcal{S}_{1,k}(u) = (-1)^{P+Q+PQ} e^{-\frac{P+Q}{2}t} u^{-PQ}
  \left( 1 + \cO (u) \right) \, ,
  & k = P \, , \\
  \mathcal{S}_{2,k}(u) = (-1)^{PQ} e^{-\frac{P+Q}{2}t} u^{-PQ}
  \left( 1 + \cO (u) \right) \, ,
  & k = P \, , \\
  \mathcal{S}_{i,k}(u) = \cO (u^{-PQ} u^{(k-P)P}) \,,
  & P+1 \le k \le P+Q-1 \, , \\
 \end{cases}
 \ \mbox{at} \quad u \to 0 \, , 
\end{align}
\begin{align}
 &
 \begin{cases}
  \mathcal{S}_{i,k}(u) = \cO (u^{kP}) \,,
  & 1 \le k \le Q-1 \, , \\
  \mathcal{S}_{1,k}(u) = (-1)^{P+Q+PQ} e^{-\frac{P+Q}{2}t} u^{PQ}
  \left( 1 + \cO (1/u) \right) \, ,
  & k = Q \, , \\
  \mathcal{S}_{2,k}(u) = (-1)^{PQ} e^{-\frac{P+Q}{2}t} u^{PQ}
  \left( 1 + \cO (1/u) \right) \, ,
  & k = Q \, , \\
  \mathcal{S}_{i,k}(u) = \cO (u^{PQ} u^{-(k-Q)Q}) \,,
  & Q+1 \le k \le P+Q-1 \, , \\
 \end{cases}
 \ \mbox{at} \quad u \to \infty \, .
\end{align}
Thus we obtain that $ \mathcal{S}(u,f)$ is a polynomial of $f$ and of $u^{PQ}$
\begin{align}
 \mathcal{S}(u,f)
  & =
 f^{2(P+Q)} + 1 +
 (-1)^{P+Q} e^{-(P+Q)t} \left( u^{-2PQ} f^{2Q} + u^{2PQ} f^{2P} \right)
  + \cdots \, .\label{SNewtpol}
\end{align}
where $\cdots$ means terms within the Newton polygon, and which are not determined by asymptotic behaviors.
The spectral curve for the two-cut matrix model is now given by the
following polynomial relation,
\begin{equation}
 \mathcal{S}(u,f)
  = 0 \, .
  \label{sp_curve_saddle_pt}
\end{equation}

It is shown in \cite{Borot:2013lpa,Borot:2013pda} 
that if the potential has convexity property, then the solution of the saddle point analysis is unique. This is the case here.
We are looking for a two-cut solution, i.e. a genus-one spectral curve.
A generic ${\cal S}(u,f)$ with the given Newton's polygon (\ref{SNewtpol}), would have genus the number of interior points of the Newton's polygon, i.e. $4(P+Q)-3$.
However we are here looking for a genus-one curve, which implies that all but one coefficient of ${\cal S}(u,f)$ must be fixed by vanishing of some discriminants.
The last coefficient is fixed by the filling fraction condition for
$\U(N|M)$, corresponding to (\ref{FF01}) and (\ref{FF02}):
\begin{align}
  \frac{1}{2\pi i} \oint_{{\cal A}} W(u) \, du 
 = g_s N \, , \qquad
  \frac{1}{2\pi i} \oint_{{\cal B}} W(u) \, du 
 = g_s M \, , 
\end{align}
where $W(u)$ is the sum of the resolvents, and $\mathcal{A}$ and $\mathcal{B}$ are contours around the first cut and the second cut.
One could determine ${\cal S}(u,f)$ by solving all vanishing discriminant equations, so as to impose that the curve have genus-one.
However there is a short-cut: one can exhibit an ${\cal S}(u,f)$ with
the correct Newton's polygon, the correct filling fraction condition
and the correct asymptotic behaviors at $0$ and $\infty$ and which is
guaranteed to be genus-one. By unicity according to
\cite{Borot:2013lpa,Borot:2013pda} 
it must be the correct spectral curve.

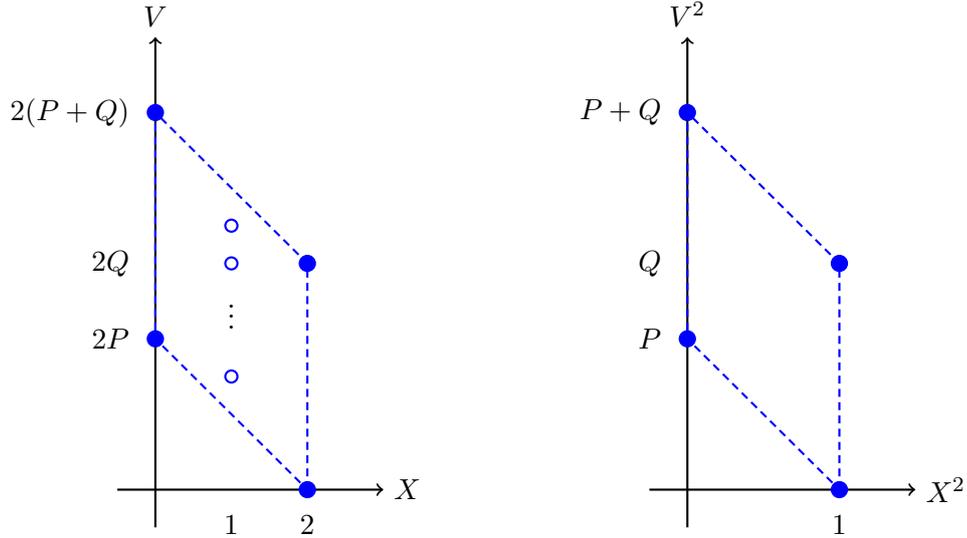
\begin{figure}[t]
 \begin{center}
  \begin{tikzpicture}[thick]
   
   \draw [->] (-0.5,0) -- (3,0) node [right] {$X$};
   \draw [->] (0,-0.5) -- (0,6) node [above] {$V$};

   \draw [->] (7-0.5,0) -- (10,0) node [right] {$X^2$};
   \draw [->] (7,-0.5) -- (7,6) node [above] {$V^2$};

   \node (X1=1) at (1,-0.2) [below] {$1$};
   \node (X1=2) at (2,-0.2) [below] {$2$};

   \node (X2=1) at (9,-0.2) [below] {$1$};

   \node (V1=P) at (-0.2,2) [left] {$2P$};
   \node (V1=Q) at (-0.2,3) [left] {$2Q$};
   \node (V1=P+Q) at (-0.2,5) [left] {$2(P+Q)$};

   \node (V2=P) at (7-0.2,2) [left] {$P$};
   \node (V2=Q) at (7-0.2,3) [left] {$Q$};
   \node (V2=P+Q) at (7-0.2,5) [left] {$P+Q$};

   \foreach \x in {0,7} 
   {

   \draw [blue, thick,densely dashed] 
   (\x+2,0) -- (\x+2,3) -- (\x,5) -- (\x,2) -- cycle;

   \filldraw[blue] (\x+2,0) circle (0.1);
   \filldraw[blue] (\x,5) circle (0.1);
   \filldraw[blue] (\x+2,3) circle (0.1);
   \filldraw[blue] (\x,2) circle (0.1);

   }


   \draw[blue, thick] (1,3.5) circle (0.08);
   \draw[blue, thick] (1,3) circle (0.08);
   \node (ddots) at (1,2.4) {$\vdots$};
   \draw[blue, thick] (1,1.5) circle (0.08);


  \end{tikzpicture}
 \end{center}
 \caption{Newton's polygons for the spectral curve.
 The polygon in the left panel corresponds to the curve
 (\ref{S_curve_XV02}). 
 There are cetain lattice points inside the polygon, which are fixed by the condition that the curve should have genus-one, and by the filling fraction condition.
 The right panel shows its singular limit, where the two resolvents
 $W^{(1)}(u)$ and $W^{(2)}(u)$ coincide with each other.
 In this case there is no lattice point inside the polygon.
 }
 \label{N-poly}
\end{figure}

The ${\cal S}(u,f)$ which has genus-one and the correct asymptotic behavior is simply the symplectic transform of the unknot of section \ref{sec:symplectic}, which we write parametrically as~\cite{Brini:2011wi}
%
%
%
%
%
\begin{equation}
 f = \frac{V}{U} \, , \qquad
 u^{-PQ} = X = U^Q V^P \, ,
\end{equation}
we obtain another expression of the spectral curve (\ref{sp_curve_saddle_pt})
in terms of $U$ and $V$,
\begin{equation}
 \cS (U,V)
  =   
  U^{2(P+Q)} + V^{2(P+Q)} 
 + (-1)^{P+Q} e^{-(P+Q)t} (UV)^{2(P+Q)}
 + (-1)^{P+Q} e^{-(P+Q)t}
 + \cdots
 = 0 \, ,   
\end{equation}
which is equivalent to
\begin{equation}
 \cS (X,V) =
 V^{2(P+Q)} 
  + X^2 
  + (-1)^{PQ} e^{-(P+Q)t} X^2 V^{2Q}
  + (-1)^{PQ} e^{-(P+Q)t} V^{2P}
  + \cdots
  = 0 \, .
  \label{S_curve_XV02} 
\end{equation}
We depict the Newton's polygon for this spectral curve in
Fig.~\ref{N-poly}.

Let us then comment on the singular limit of the spectral curve
(\ref{S_curve_XV02}), which is realized when the two resolvents
$W^{(1)}(u)$ and $W^{(2)}(u)$ coincide with each other.
In this case the analytic function obeys $\cS(u,f) =
\cS(u\,\omega^{\frac{1}{2}}, f)$.
This implies it depends only on $u^{2PQ}$.
Therefore the spectral curve (\ref{S_curve_XV02}) has to be written
only in terms of $X^2$ and $V^2$.
The Newton's polygon corresponding to this situation is shown in the right
panel of Fig.~\ref{N-poly}.
Since there is no lattice point inside the polygon, the spectral curve
has no free parameter.
This means that it is just reduced to the genus-zero curve, which
corresponds to the limit $\zeta \to 0$ of the curve (\ref{ABJM_curve00}).

\subsection{Asymptotic expansion and topological recursion}

The matrix integral (\ref{ABJM_MM_torus}) is of the type discussed in \cite{Borot:2013lpa}, therefore it guarantees that it has an asymptotic expansion of the type
\begin{equation}
\log \cZ_{\rm ABJM}^{(P,Q)} = \sum_{g=0}^\infty g_s^{2g-2} F_g
\end{equation}
which obeys the topological recursion of \cite{Eynard:2007kz}, i.e.
$F_g$ is the $g^{\rm th}$ symplectic invariant of the spectral curve as defined by the topological recursion \cite{Eynard:2007kz}.
Similarly, all expectation values have a $g_s$-expansion, whose coefficients are given by the topological recursion.

The character expectation values $\Big<\mathrm{Tr}_R U \Big>$ can be decomposed on the basis of power sums.
We have:
\begin{equation}
 \left<
  \prod_{i=1}^n \Tr U^{p_i}
 \right> 
 = {\rm Res\,}_{x_1\to 0}\dots {\rm Res\,}_{x_n\to 0} W_n(x_1,\dots,x_n)\,\,x_1^{p_1}\dots x_n^{p_n} \prod_{i=1}^n \frac{dx_i}{x_i}
\end{equation}
and $W_n$ has a topological expansion:
\begin{equation}
W_n(x_1,\dots,x_n) = \sum_{g=0}^\infty g_s^{2g-2+n} W_{g,n}(x_1,\dots,x_n)
 \label{n-pt_func}
\end{equation}
and $W_{g,n}$ is computed by the topological recursion.

\section{Comments on related topics}\label{sec:comments}

\subsection{Topological A-model}\label{sec:TopSt}

Let us comment on the realization of the knot invariant in topological
string theory, especially in the topological A-model.
It was shown in \cite{Ooguri:1999bv} that the knot invariant for $K$ is obtained
in the topological string by adding a brane with a proper Lagrangian
submanifold of the Calabi--Yau threefold $L_K$.
This insertion of a brane gives rise to the expectation
value of the characteristic polynomial with respect to Chern--Simons theory
\begin{align}
 \cZ_{\rm top} (K;x)
  & =
  \Big\langle
  \det \left(
	1 \otimes 1 - U \otimes e^{-x}
       \right)
  \Big\rangle_{\rm CS}
 \nonumber \\
 & =
 \sum_{n=0}^\infty 
 \Big\langle
 \mathrm{Tr}_{R_n} U
 \Big\rangle_{\rm CS} \,
 e^{-nx}
  \, ,
 \label{top_part_func01}
\end{align}
where the matrix $U$ is the holonomy along the knot $K$, 
$\displaystyle U = \mathrm{P} \exp \left( \oint_K A\right)$, and $R_n$
is the totally symmetric representation with $n$ boxes.
This means that this topological string partition function is the
discrete Fourier (Laplace) transform of the HOMFLY polynomial, since the
expectation value of the holonomy $\Big\langle\mathrm{Tr}_{R_n}
U\Big\rangle$ just gives the knot invariant.
If we consider a multi-point correlator of the characteristic
polynomials, the knot invariant with more generic representations is obtained.
Note that this knot invariant is analogous to the matrix integral with
the external source, as discussed in Sec.~\ref{sec:MM}.
This kind of relation between the characteristic polynomial and the
external source is naturally interpreted from the viewpoint of the
topological expansion of spectral curves.

In our case it is natural to consider a supermatrix version of the
partition function (\ref{top_part_func01}), corresponding to ABJM theory
\begin{equation}
 \cZ_{\rm top} (K;x,y)
  =
  \left\langle
  \Sdet \left(
	 1 \otimes 1 - U \otimes 
	 \left(\begin{array}{cc} e^{-x} & \\ & e^{-y} \end{array}\right)
	\right)
  \right\rangle_{\rm ABJM}
  \, .
 \label{top_part_func02}
\end{equation}
Actually one can obtain this partition function by applying both of
bosonic and fermionic modes, describing strings stretching between the three-sphere $S^3$ and the Lagrangian $L_K$
\begin{align}
 \cZ_\text{top} (K;x,y)
 & =
 \left<
 \frac{\det \left(1 \otimes 1 - U' \otimes e^{-x}\right)}
      {\det \left(1 \otimes 1 - U'' \otimes e^{-y}\right)}
 \right>
\end{align}
where $U = \operatorname{diag}(U',U'')$.
Whether it comes in denominator (bosonic) or numerator (fermionic) depends on the relative orientation of $S^3$ and $L_K$~\cite{Aganagic:2003db}.
See also the argument on the characteristic polynomial ratio presented in Sec.~\ref{sec:MM}.

We can similarly expand (\ref{top_part_func02}) with the corresponding
expectation value $\Big\langle \mathrm{Str}_{R_n} U \Big\rangle_{\rm ABJM}$.
This partition function is also expanded with the string coupling
constant in the sense of the WKB expansion~\cite{Ooguri:1999bv}.
Using the identity
\begin{equation}
 \Sdet \left(
	1 \otimes 1 - U \otimes 
	\left(\begin{array}{cc} e^{-x} & \\ & e^{-y} \end{array}\right)
       \right)
 = 
 \exp \left[
       \Str \log \left(
		  1 \otimes 1 - U \otimes 
  		\left(\begin{array}{cc} e^{-x} & \\ & e^{-y} \end{array}\right)
		 \right)
      \right]
 \, ,
\end{equation}
we have
\begin{equation}
 \cZ_{\rm top}(K;x,y)
  \ \sim \
  \exp \left(
	\frac{1}{g_s} \int_y^x p(x) dx
       \right)
  \, ,
  \label{disc_amp}
\end{equation}
where the integrand is given by
\begin{equation}
 p(x) 
  = 
  \lim_{g_s \to 0}
  \sum_{n=0}^\infty g_s
  \Big\langle \mathrm{Str} \, U^n \Big\rangle_{\rm ABJM}
  \, e^{-nx}
  \, .
  \label{one-form}
\end{equation}
In this way we can show that its leading contribution is given by the
disc amplitude as well as the ordinary knot invariant.
In this case, since the ABJM matrix model is obtained from
Chern--Simons theory on the lens space $L(2,1)$ through the analytic
continuation, the mirror curve, on which the one-form is defined, is
replaced with that for the local $\mathbb{P}^1 \times \mathbb{P}^1$ geometry.
Actually the Wilson loop expectation value is evaluated based on this
spectral curve~\cite{Marino:2009jd,Klemm:2012ii}.
Furthermore, in (\ref{disc_amp}), both of the initial and end points of the
integral have physical meanings, as positions of brane and
anti-brane.
Thus the partition function describes pair creation of branes in the
topological string.
If we take the limit $y \to \infty$, it goes back to the usual one,
including either of bosonic or fermionic modes.

\subsection{Topological B-model}

In the B-model description of the topological strings, 
the $n$-point function defined in (\ref{n-pt_func}) plays an important role.
As well as the one-point function $W_1=W$ which is given as the resolvent, we
also have a series expansion of the multi-point function with respect to
the coupling constant.
If once a spectral curve is obtained, one can determine higher order
terms by using the topological recursion~\cite{Eynard:2007kz}.
In this study the spectral curve is given by the genus-one mirror curve
for the local $\bbP^1 \times \bbP^1$ geometry and its symplectic transform.
Indeed this expansion has a natural interpretation in terms of the
B-model topological strings.
The multi-point correlation function corresponds to multiple insertion
of the Wilson loop operators into the Chern--Simons matrix model.
Thus a set of variables in the multi-point function (\ref{n-pt_func})
provides the boundary condition for topological strings.
This implies that it computes the open string sector of the B-model, and
we can obtain the corresponding open Gromov--Witten
invariants based on the mirror symmetry in a perturbative
way~\cite{Marino:2006hs,Bouchard:2007ys,Eynard:2012nj}.

As in the case of the topological A-model, the knot invariant can be 
investigated also in the B-model through the mirror symmetry.
In addition to the unknot
invariants~\cite{Aganagic:2000gs,Aganagic:2001nx}, the torus knot
invariant is formulated by using the symplectic transformation acting on
the B-model open string moduli~\cite{Brini:2011wi}.%
\footnote{
Let us note that another kind of approach to the B-model description,
which is in principle applicable to any knots, is discussed based on the
A-polynomial~\cite{Aganagic:2012jb}.}
Moreover this prescription for giving the torus knot invariant is now
applied to the knot invariant not only in the three-sphere $S^3$, but
also in the lens space $L(2,1)$~\cite{Jockers:2012pz,Stevan:2013tha}.
Since the ABJM matrix model is perturbatively equivalent to 
Chern--Simons theory on the lens space $L(2,1)$, the perturbative
analysis of the knot invariant for $L(2,1)$, which is based on the
topological recursion, is apparently relevant to our supermatrix model
for the torus knot introduced in Sec.~\ref{sec:torus}.
Although we can obtain a systematic expansion of the correlation
functions, we have to also take into account non-perturbative
contributions, which play an important role in knot
theory~\cite{Dijkgraaf:2010ur,Borot:2012cw} and ABJM
theory~\cite{Hatsuda:2013oxa}.
The study of such a non-perturbative effect on the torus knots is an
interesting and important issue to be clarified in the future.

\subsection{Matrix models} \label{sec:MM}

We comment on the idea underlying this work, which comes from random matrices.
In random matrix theory, the wave functions are expectation values of
characteristic polynomials
\begin{equation}
\psi(x) = \Big\langle \det(x-M) \Big\rangle \, ,
\end{equation}
where the expectation value is taken with respect to a certain measure, which is specified later.
The Hamiltonian associated with this wave function is given by the
non-commutative Riemann surface, which is obtained through quantization of
the spectral curve.
In fact, a more important object is the kernel
\begin{equation}
 \cK(x;y) = \frac{1}{y-x} \left< \frac{\det(x-M)}{\det(y-M)}\right> \, .
\end{equation}
Out of that kernel one can reconstruct every other observables.
For instance, if the matrix size is given by $N$, the wave function is
obtained by sending $y\to\infty$,
\begin{equation}
 \psi(x) = \lim_{y\to\infty} y^{N+1} \,\cK(x;y) \, .
\end{equation}
Every other correlations of characteristic polynomials is obtained by
the Fay identity~\cite{Fay:1973} (See
also~\cite{Borodin:2006AAM,Bergere:2009zm}) 
\begin{equation}
 \left< \prod_{i=1}^k \frac{\det(x_i-M)}{\det(y_i-M)} \right>
  = \frac{\prod_{i,j}(y_i-x_j)}{\prod_{i<j}(x_i-x_j)\,(y_i-y_j)}\,\, 
  \det_{1\leq i,j\leq k} \cK(x_i;y_j) \, ,
\end{equation}
which is also seen as Pl\"ucker or Hirota equation.
The factor in front of the determinant of the kernel can be written as
the Cauchy determinant (\ref{Cauchy_det}).


Wave functions and kernels can be seen themselves as partition functions. Indeed let
\begin{equation}
 \cZ=\int d\mu(M)
\end{equation}
be a matrix integral with some measure $d\mu(M)$ depending on a coupling constant $g_s$.
In many cases, there is a small $g_s$ expansion of the type
\begin{equation}
 \log \cZ = \sum_{g=0}^{\infty} g_s^{2g-2}\,F_g(\cC)
\end{equation}
where $\cC$ is the spectral curve associated to the measure $d\mu$, i.e. the $g_s\to 0$ limit of the eigenvalue distribution.

An expectation value of characteristic polynomials can be written
\begin{equation}
 \cK(x_1,\dots,x_k;y_1,\dots,y_k)
 =
 \frac{1}{\cZ}
 \int d\mu(M)\prod_{i=1}^k \frac{\det(x_i-M)}{\det(y_i-M)} \, ,
\end{equation}
i.e. it is the partition function with a new measure
\begin{equation}
d\mu_{\bm x;\bm y}(M) 
 = 
 \exp {\sum_{i=1}^k \Big( \Tr \log(x_i-M)- \Tr \log(y_i-M) \Big)} \, 
 d\mu(M) \, .
\end{equation}
There is also a $g_s$ expansion with a new spectral curve
\begin{equation}
 \log \cK(x_1,\dots,x_k;y_1,\dots,y_k) 
 = 
 \sum_{g=0}^{\infty} g_s^{2g-2}\,(F_g(\cC_{\bm x;\bm y})-F_g(\cC)) \, .
\end{equation}
The way to find the new spectral curve is as follows. 
The spectral curve $\cC$ is a Riemann surface equipped with two
analytic functions $u$ and $v$ as $\cC = \{ (u,v) \in \C \times \C \, |
\, H(u,v) = 0 \}$.
The new spectral curve $\cC_{\bm x;\bm y}$ is a Riemann surface such
that $vdu$ has additional simple poles at the $x_i$'s (residue $+1$) and
at the $y_i$'s (residue $-1$).


Instead of characteristic polynomials one may also be interested in external fields interactions
\begin{equation}
\left< {\rm e}^{\Tr M A} \right> 
 = 
 \frac{1}{\cZ}\int d\tilde\mu(M)\,\,{\rm e}^{\Tr M A} \,.
\end{equation}
Again, this has (under some assumptions) a $g_s$ expansion with a new spectral curve $\cC_A$
\begin{equation}
 \log \left<{\rm e}^{\Tr M A} \right> 
 = \sum_{g=0}^{\infty} g_s^{2g-2}\, 
 \left(
  F_g(\cC_{A}) - F_g(\cC) 
 \right) 
 \, .
\end{equation}
The way to find the new spectral curve is as follows. The spectral curve $\cC$ is a Riemann surface equipped with two analytic functions $u$ and $v$.
The new spectral curve $\cC_{A}$ is a Riemann surface such that
$udv$ has additional simple poles at the $a_i$'s (the eigenvalues of $A$) with a residue equal to the multiplicity of $a_i$.

We see that under the exchange $u \leftrightarrow v$ and by identifying the $a_i$'s with the $x_i$'s (multiplicity $\alpha_i=1$) and the $y_i$'s (multiplicity $\alpha_i=-1$), this would be the same spectral curve and thus
\begin{equation}
 \left< {\rm e}^{\Tr M A} \right>_{\tilde\mu}
 \propto 
 \left< \prod_{i=1}^k \det(x_i-M)^{\alpha_i} \right>_{\mu}
\end{equation}
where the measures $\mu$ and $\tilde\mu$ are obtained by exchanging the
role of $u$ and $v$ in the spectral curves, and $A$ is a matrix with
eigenvalues $x_i$ with multiplicity
$\alpha_i$~\cite{Eynard:2007kz,Eynard:2007nq,Eynard:2013csa,Kimura:2014mua,Kimura:2014vra}.
This raises the question of how can a multiplicity be negative?
This is why supermatrix models are needed. Negative multiplicities correspond to the fermionic side of the supermatrix \cite{Desrosiers:2009pz}. This duality between external field and expectation values of characteristic polynomials appeared in \cite{Brezin:2000CMP}.


As shown in (\ref{ABJM_WL_ext}), the half-BPS Wilson loop expectation
value in ABJM theory has a quite similar expression to that of matrix
integral with the external source.
The Gaussian matrix model with the external source is given as follows,
\begin{align}
 \left\langle e^{\Tr MA} \right\rangle
  & =
 \frac{1}{\cZ}
  \int dM \, e^{-\frac{1}{2g_s} \Tr M^2 + \Tr M A}
 \nonumber \\
  & =
 \frac{1}{\cZ}
  \frac{1}{\Delta(a)}
  \int \prod_{i=1}^N \frac{dx_i}{2\pi} \,
  e^{-\frac{1}{2g_s} x_i^2}
 \det_{1 \le i, j \le N} e^{x_i a_j}
 \prod_{i<j}^N (x_i - x_j)
 \, ,
\end{align}
where the $\U(N)$ part is integrated out using the
Harish-Chandra--Itzykson--Zuber formula~\cite{HC:1957,Itzykson:1979fi}.
Writing the $\U(N)$ character in terms of the Schur function
\begin{equation}
 s_\lambda (e^x) 
  =
  \frac{1}{\Delta(e^x)} 
  \det_{1 \le i, j \le N} e^{x_i \left(\lambda_j+N-j\right)}
  \, ,
\end{equation}
we obtain the Wilson loop expectation value with respect to the $\U(N)$
Chern--Simons theory
\begin{equation}
 \Big\langle
  W_R(\unknot)
 \Big\rangle_{\U(N)}
 =
 \frac{1}{\cZ_{\rm CS}} 
 \int \prod_{i=1}^N \frac{dx_i}{2\pi} \,
 e^{-\frac{1}{2g_s} x_i^2}
 \det_{1 \le i, j \le N}
 e^{x_i \tilde{\xi}_j}
 \prod_{i<j}^N \left( 2 \sinh \frac{x_i-x_j}{2} \right)
 \, ,
 \label{CS_ext_source}
\end{equation}
with $\tilde{\xi}_i = \lambda_i - i + \frac{N+1}{2}$.
In this sense insertion of the Wilson loop operator corresponds to
the external fields for the matrix integral.

This kind of interpretation can be possible even
for the supergroup character average (\ref{ABJM_WL_ext}), at least as
long as the partition $\lambda$, corresponding to the representation
$R$, satisfies $d(\lambda)=N$.%
\footnote{When the representation does not satisfy
this condition, there is no simple analogy between the matrix integral
with the external fields and the Wilson loop average (\ref{WL_hook_rep}),
because such an external field has to consist of $N+N$ parameters for
$\U(N|N)$ theory.
On the other hand, this analogy holds for arbitrary representations in
the ordinary $\U(N)$ Chern--Simons theory as (\ref{CS_ext_source}).
It is because even if the number of non-zero elements in the partition
is less than $N$ at the first place, it can be made $N$ by the constant
shift, since the average (\ref{unknot_CS}) is invariant under the shift
of the partition
\begin{equation}
 (\lambda_1, \lambda_2, \cdots, \lambda_N) 
  \ \longrightarrow \
  (\lambda_1 + c, \lambda_2 + c,\cdots, \lambda_N + c) 
 \, .
\end{equation}
It is obvious that the $\U(N|N)$ invariant (\ref{ABJM_WL_ext}) is not
invariant under such a constant shift of the partition, since the
corresponding external field is characterized by the Frobenius
coordinates of the partition.
}
%
%
All this means that characters of $\U(N|N)$ 
are dual to expectation values of characteristic polynomials of another
matrix model with some measure $\mu$,
\begin{equation}
\Big< W_R(K) \Big>_{\U(N|N)} 
 = \left<\prod_{i=1}^N \frac{\det(\alpha_i-M)}{\det(\beta_i-M)}\right>_\mu
 \, ,
\end{equation}
and where $\alpha_i, \beta_i$ are the Frobenius coordinates for the
representation $R$.
Then the Fay identity says that
\begin{equation}
 \Big< W_R(K) \Big>_{\U(N|N)} 
 = \det_{1 \le i, j \le N} \Big< W_{(\alpha_i|\beta_j)}(K) \Big>_{\U(1|1)} \, .
\end{equation}
This is what we have checked in this article, especially as the
determinantal formula for the unknot average shown in (\ref{unknot_NN01}).
This means that those supergroup character expectation
values are Giambelli compatible~\cite{Borodin:2006AAM}.

\section{Discussion}\label{sec:summary}

In this paper we have considered the supergroup character expectation
value based on the ABJM matrix model as the supermatrix Chern--Simons
theory, with emphasis on its connection to the knot invariant.
We have explicitly computed the $\U(1|1)$ expectation value for the unknot and
torus knot matrix models.
We have obtained the determinantal formula, where the $\U(1|1)$ average
plays a role of a building block for $\U(N|N)$ theory, and shown the
Rosso--Jones-type formula for the supergroup character average.
We have also discussed the $\U(N|M)$ theory, and found its
determinantal formula which interpolates $\U(N|N)$ and $\U(N)$ theories.
We have derived the spectral curve for the torus knot as the
symplectic transform of the unknot curve, and by analyzing the saddle
point equations for the torus knot matrix model itself.
We have shown these two methods to obtain the spectral curve are
consistent with each other.
We have then commented on how to realize the knot
invariant in topological string theory, and the underlying idea coming
from random matrix theory.

Let us comment on some open issues to be investigated in the future.
The most attractive one is to check whether the supergroup character
average can be a knot invariant or not.
We have certain evidence on this point.
First of all, the $\U(N|N)$ character expectation value contains $\U(N)$ part as shown for the unknot Wilson loop (\ref{unknot_NN02}).
Furthermore, in the limit $M \to 0$, the $\U(N|M)$ matrix model \eqref{ABJ_MM_torus} is explicitly reduced to the $\U(N)$ torus knot matrix model, which is proven to be a topological invariant of the torus knot.
This implies that, at least for the torus knot, the $\U(N|M)$ average would play a role of the knot invariant which contains much information than the ordinary $\U(N)$ invariant.

Secondly the ABJM matrix model is obtained from the lens space
Chern--Simons theory through the analytic continuation.
Since the knot invariant in the lens space is given as the character
average with the corresponding matrix model, we can expect the character
expectation value with the supermatrix model is also a knot invariant,
at least up to some non-perturbative contributions.
Moreover, the construction of the knot invariant shown in
Sec.~\ref{sec:TopSt} seems natural from the viewpoint of topological
string, and possibly applied to an arbitrary knot.
These supporting facts suggest that we can obtain a knot invariant from
ABJM theory.
For this purpose, for example, it is interesting to see whether the torus
knot supermatrix model can be directly obtained from ABJM theory.
In the case of classical group theory, it is shown by~\cite{Tanaka:2012nr}
that the torus knot matrix model is obtained from $\N=2$ Chern--Simons
theory on the ellipsoid-type squashed three-sphere $S^3_b$.
This suggests that the supergroup torus knot character average is given by the half-BPS Wilson loop for the $\N=6$ ABJM theory on $S^3_b$.
However the integral formula (\ref{ABJM_MM_torus}) is not obtained
naively using the localization method for $S^3_b$, because the
matter contribution cannot be written as a simple $\cosh$ function for such a case~\cite{Hama:2011ea}.
In addition, the mirror description of the torus knot model \eqref{ABJM_MM_torus}, presented in \eqref{ABJM_mirror_torus}, is different from that on $S^3_b$.
Although it is not yet obvious that such a way to obtain the torus knot invariant using the ellipsoid $S^3_b$ is applicable to the present case, this discrepancy would reflect that ABJM theory itself is not topological, and we may need to consider, for example, topological twist to obtain the knot invariant for the Seifert manifold~\cite{Lawrence:1999CMP,Beasley:2009mb,Kallen:2011ny,Brini:2011wi}.
We will return to this issue in the future.

If we can have the supergroup knot invariant, it is also interesting to
provide another definition of the $\U(N|N)$ knot invariant, for example,
based on the Skein relation, which enable us to compute the invariant
for any knots in principle.
It is naively expected that such a relation can be related to the
supergroup WZW model.
However, it is known that the $\U(1|1)$ WZW model describes the Skein
relation for the Alexander--Conway
polynomial~\cite{Deguchi:1989zu,Kauffman:1990ix,Rozansky:1992td}, while 
the torus knot average obtained in this paper (\ref{torus_U(11)}) is
not consistent with that.
This reflects the fact that ABJM theory is not just Chern--Simons
theory with the supergroup.
We have to explore other possibility of the two-dimensional
conformal field theory model, which shall live on the boundary of ABJM theory, if it exists.
Furthermore, as the recent progress in knot theory, it is definitely
interesting to study the volume conjecture and its
generalization~\cite{Kashaev:1996kc,Murakami:2001AM,Gukov:2003na}, the AJ
conjecture~\cite{Garoufalidis:2004GTM}, and also the knot homology~\cite{Khovanov:2000DMJ,Khovanov:2008FM,Khovanov:2008GT,Gukov:2004hz,Dunfield:2005si}
corresponding to the supergroup knot invariant.
In order to discuss the volume conjecture, we have to deal with the
hyperbolic knot and the volume of its complement in $S^3$.
Therefore our construction, which is so far available only for
the unknot and torus knots, is not yet enough to study this conjecture.
Also from this point of view, the definition of the knot invariant based
on the Skein relation is highly desirable, as commented above.
The AJ conjecture claims that the knot invariant satisfies some
integrable equations, which are obtained through quantization of the
A-polynomial.
In other words, this implies that the knot invariant is associated with
the corresponding $\tau$-function.
As pointed out in Sec.~\ref{sec:comments}, the knot invariant is closely
related to the matrix integral with the external fields, which can be
seen as a certain $\tau$-function.
For example, the determinantal formula for the character expectation
value given in this paper can be one of the supporting results for such
a suggestive relation.
The determinantal formula also gives an insight into the knot homology.
Using the Jacobi identity for determinants, one can obtain some
relations between the knot invariants for different rank groups.
Such a relation could provide a natural differential on the knot homology.

\subsection*{Acknowledgements}

We would like to thank G. Borot and M. Mari\~no for fruitful discussions.
We are also grateful to A. Brini for carefully reading the manuscript
and giving useful comments.
BE thanks Centre de Recherches Math\'ematiques de Montr\'eal, 
the FQRNT grant from the Qu\'ebec government, 
Piotr Su\l kowski and the ERC starting grant Fields-Knots.
The work of TK is supported in part by Keio Gijuku Academic Development Funds, MEXT-Supported Program for the Strategic Research Foundation at Private Universities ``Topological Science'' (No.~S1511006), and JSPS Grant-in-Aid for Scientific Research on Innovative Areas ``Topological Materials Science'' (No.~JP15H05855).


\appendix


\section{Mirror of the torus knot ABJM partition function}
\label{sec:mirror}

In this appendix we derive the mirror description of the torus knot
ABJM partition function~\cite{Kapustin:2010xq}.
We first expand the determinants in (\ref{ABJM_MM_torus}) as summation
over permutations
\begin{equation}
 \cZ_{\rm ABJM}^{(P,Q)}
  =
  \sum_{\sigma,\sigma' \in \mathfrak{S}_N} (-1)^{\sigma+\sigma'}
  \frac{1}{N!^2} 
  \int [dx]^N [dy]^N
  \prod_{i=1}^N
  \left(
   2 \cosh \frac{x_i - y_{\sigma(i)}}{2P} \,
   2 \cosh \frac{x_i - y_{\sigma'(i)}}{2Q}
  \right)^{-1}
  \, .
\end{equation}
Applying the formula (\ref{cosh_FT}), we have
\begin{eqnarray}
 &&
  \sum_{\sigma,\sigma' \in \mathfrak{S}_N} (-1)^{\sigma+\sigma'}
  \frac{1}{N!^2} 
  \int \!
  \frac{d^N x}{(2\pi)^N} \frac{d^N y}{(2\pi)^N}
  \frac{d^N z}{(2\pi)^N} \frac{d^N w}{(2\pi)^N} \,
  \prod_{i=1}^N
  \left(
   \cosh z_i \, \cosh w_i
  \right)^{-1}  
  \nonumber \\
 && 
  \hspace{2em} 
  \times
  \exp \left[
	\frac{ik}{4PQ\pi} \sum_{i=1}^N (x_i^2 - y_i^2)
	+ \frac{i}{\pi} \sum_{i=1}^N
	\left(
	 x_i \left( \frac{z_i}{P} + \frac{w_i}{Q} \right)
	 - y_i \left( \frac{z_{\sigma^{-1}(i)}}{P} 
		+ \frac{w_{\sigma'^{-1}(i)}}{Q} \right)
	\right)
       \right]
  \nonumber \\
  & = &
  \sum_{\sigma,\sigma' \in \mathfrak{S}_N} (-1)^{\sigma+\sigma'}
  \frac{(PQ)^N}{N!^2 \, k^N} 
  \int \!
  \frac{d^N z}{(2\pi)^N} \frac{d^N w}{(2\pi)^N} \,
  \prod_{i=1}^N
  \frac{\exp \left[
	      - \frac{2i}{k\pi} 
	      \left( z_i w_i - z_{\sigma^{-1}(i)} w_{\sigma'^{-1}(i)} \right)
	     \right]}
       { \cosh z_i \, \cosh w_i }
 \, .       
\end{eqnarray}
At this moment it is obvious that the partition function depends only on the
composition of permutations $\sigma \cdot \sigma'^{-1}$.
Thus, by fixing either of them $\sigma'$ as the trivial permutation, we obtain
\begin{align}
 \cZ_{\rm ABJM}^{(P,Q)}
  & = 
  \sum_{\sigma \in \mathfrak{S}_N} (-1)^N
  \frac{(PQ)^N}{N! \, k^N}
  \int \!
  \frac{d^N z}{(2\pi)^N} \frac{d^N w}{(2\pi)^N} \,
  \prod_{i=1}^N
  \frac{\exp \left[
	      - \frac{2i}{k\pi} 
	      \left( z_i - z_{\sigma(i)} \right) w_i
	     \right]}
       { \cosh z_i \, \cosh w_i }
 \nonumber \\
 & = 
  \sum_{\sigma \in \mathfrak{S}_N} (-1)^N
  \frac{(PQ)^N}{N! \, k^N}
  \int \!
  \frac{d^N z}{(2\pi)^N} \,
  \prod_{i=1}^N
  \left(
   \cosh z_i \cdot 2\cosh \frac{z_i - z_{\sigma(i)}}{k}
  \right)^{-1}
 \nonumber \\
 & = 
  \frac{(PQ)^N}{N! (2k)^N}
  \int \! \frac{d^N z}{(2\pi)^N} \,
  \prod_{i<j}^N \left( \tanh \frac{z_i - z_j}{2k} \right)^2 \,
  \prod_{i=1}^N \left( 2 \cosh \frac{z_i}{2} \right)^{-1}
  \nonumber \\
 & = 
  (PQ)^N \cZ_{\rm ABJM}^{(1,1)}
  \, .
  \label{ABJM_mirror_torus}
\end{align}
This is the mirror expression of the partition function
(\ref{ABJM_MM_torus}). 
Especially for $k=1$, the mirror theory turns out to be $\N=4$
SYM theory with a single fundamental and a single adjoint
hypermultiplet.
The dependence on the parameters $(P,Q)$ becomes trivial in this mirror
representation.


\bibliographystyle{ytphys}
\bibliography{abjm_knot}

\providecommand{\href}[2]{#2}\begingroup\raggedright\begin{thebibliography}{10}

\bibitem{Witten:1988hf}
E.~Witten, ``{Quantum field theory and the Jones polynomial},''
\href{http://dx.doi.org/10.1007/BF01217730}{{\em Commun. Math. Phys.}
  {\bfseries 121} (1989) 351--399}.

\bibitem{Lawrence:1999CMP}
R.~Lawrence and L.~Rozansky, ``{Witten--Reshetikhin--Turaev Invariants of
  Seifert Manifolds},'' \href{http://dx.doi.org/10.1007/s002200050678}{{\em
  Commun. Math. Phys.} {\bfseries 205} (1999) 287--314}.

\bibitem{Beasley:2009mb}
C.~Beasley, ``{Localization for Wilson Loops in Chern--Simons Theory},''
  \href{http://dx.doi.org/10.4310/ATMP.2013.v17.n1.a1}{{\em Adv. Theor. Math.
  Phys.} {\bfseries 17} (2013) 1--240},
\href{http://arxiv.org/abs/0911.2687}{{\ttfamily arXiv:0911.2687 [hep-th]}}.

\bibitem{Kallen:2011ny}
J.~K\"all\'en, ``{Cohomological localization of Chern--Simons theory},''
  \href{http://dx.doi.org/10.1007/JHEP08(2011)008}{{\em JHEP} {\bfseries 1108}
  (2011) 008},
\href{http://arxiv.org/abs/1104.5353}{{\ttfamily arXiv:1104.5353 [hep-th]}}.

\bibitem{Brini:2011wi}
A.~Brini, B.~Eynard, and M.~Mari\~no, ``{Torus knots and mirror symmetry},''
  \href{http://dx.doi.org/10.1007/s00023-012-0171-2}{{\em Ann. Henri
  Poincar\'e} {\bfseries 13} (2012) 1873--1910},
\href{http://arxiv.org/abs/1105.2012}{{\ttfamily arXiv:1105.2012 [hep-th]}}.

\bibitem{Kapustin:2009kz}
A.~Kapustin, B.~Willett, and I.~Yaakov, ``{Exact Results for Wilson Loops in
  Superconformal Chern--Simons Theories with Matter},''
  \href{http://dx.doi.org/10.1007/JHEP03(2010)089}{{\em JHEP} {\bfseries 1003}
  (2010) 089},
\href{http://arxiv.org/abs/0909.4559}{{\ttfamily arXiv:0909.4559 [hep-th]}}.

\bibitem{Aharony:2008ug}
O.~Aharony, O.~Bergman, D.~L. Jafferis, and J.~Maldacena, ``{$\mathcal{N}=6$
  superconformal Chern--Simons--matter theories, M2-branes and their gravity
  duals},'' \href{http://dx.doi.org/10.1088/1126-6708/2008/10/091}{{\em JHEP}
  {\bfseries 0810} (2008) 091},
\href{http://arxiv.org/abs/0806.1218}{{\ttfamily arXiv:0806.1218 [hep-th]}}.

\bibitem{Drukker:2009hy}
N.~Drukker and D.~Trancanelli, ``{A Supermatrix model for $\mathcal{N}=6$ super
  Chern--Simons--matter theory},''
  \href{http://dx.doi.org/10.1007/JHEP02(2010)058}{{\em JHEP} {\bfseries 1002}
  (2010) 058},
\href{http://arxiv.org/abs/0912.3006}{{\ttfamily arXiv:0912.3006 [hep-th]}}.

\bibitem{Bars:1982ps}
I.~Bars, ``{Supergroups and Their Representations},''
\href{http://inspirehep.net/record/181673/}{{\em Lectures Appl. Math.}
  {\bfseries 21} (1983) 17}.

\bibitem{Berele:1987yi}
A.~Berele and A.~Regev, ``{Hook Young-diagrams with applications to
  combinatorics and to representations of Lie superalgebras},''
\href{http://dx.doi.org/10.1016/0001-8708(87)90007-7}{{\em Adv. Math.}
  {\bfseries 64} (1987) 118--175}.

\bibitem{Fay:1973}
J.~D. Fay, \href{http://dx.doi.org/10.1007/BFb0060090}{{\em {Theta Functions on
  Riemann Surfaces}}}, vol.~352 of {\em Lecture Notes in Mathematics}.
\newblock Springer Berlin Heidelberg, 1973.

\bibitem{Borodin:2006AAM}
A.~Borodin, G.~Olshanski, and E.~Strahov, ``{Giambelli compatible point
  processes},''
  \href{http://dx.doi.org/http://dx.doi.org/10.1016/j.aam.2005.08.005}{{\em
  Adv. Appl. Math.} {\bfseries 37} (2006) 209--248},
  \href{http://arxiv.org/abs/math-ph/0505021}{{\ttfamily
  arXiv:math-ph/0505021}}.

\bibitem{Bergere:2009zm}
M.~Berg\`ere and B.~Eynard, ``{Determinantal formulae and loop equations},''
\href{http://arxiv.org/abs/0901.3273}{{\ttfamily arXiv:0901.3273 [math-ph]}}.

\bibitem{Eynard:2007kz}
B.~Eynard and N.~Orantin, ``{Invariants of algebraic curves and topological
  expansion},'' \href{http://dx.doi.org/10.4310/CNTP.2007.v1.n2.a4}{{\em
  Commun. Num. Theor. Phys.} {\bfseries 1} (2007) 347--452},
\href{http://arxiv.org/abs/math-ph/0702045}{{\ttfamily arXiv:math-ph/0702045
  [math-ph]}}.

\bibitem{Dijkgraaf:2010ur}
R.~Dijkgraaf, H.~Fuji, and M.~Manabe, ``{The Volume Conjecture, Perturbative
  Knot Invariants, and Recursion Relations for Topological Strings},''
  \href{http://dx.doi.org/10.1016/j.nuclphysb.2011.03.014}{{\em Nucl. Phys.}
  {\bfseries B849} (2011) 166--211},
\href{http://arxiv.org/abs/1010.4542}{{\ttfamily arXiv:1010.4542 [hep-th]}}.

\bibitem{Borot:2012cw}
G.~Borot and B.~Eynard, ``{All-order asymptotics of hyperbolic knot invariants
  from non-perturbative topological recursion of A-polynomials},''
  \href{http://dx.doi.org/10.4171/QT/60}{{\em Quantum Top.} {\bfseries 6}
  (2015) 39--138},
\href{http://arxiv.org/abs/1205.2261}{{\ttfamily arXiv:1205.2261 [math-ph]}}.

\bibitem{Gukov:2012jx}
S.~Gukov and I.~Saberi, ``{Lectures on Knot Homology and Quantum Curves},''
\href{http://arxiv.org/abs/1211.6075}{{\ttfamily arXiv:1211.6075 [hep-th]}}.

\bibitem{Gopakumar:1998ki}
R.~Gopakumar and C.~Vafa, ``{On the gauge theory/geometry correspondence},''
  {\em Adv. Theor. Math. Phys.} {\bfseries 3} (1999) 1415--1443,
\href{http://arxiv.org/abs/hep-th/9811131}{{\ttfamily arXiv:hep-th/9811131
  [hep-th]}}.

\bibitem{Ooguri:1999bv}
H.~Ooguri and C.~Vafa, ``{Knot invariants and topological strings},''
  \href{http://dx.doi.org/10.1016/S0550-3213(00)00118-8}{{\em Nucl. Phys.}
  {\bfseries B577} (2000) 419--438},
\href{http://arxiv.org/abs/hep-th/9912123}{{\ttfamily arXiv:hep-th/9912123
  [hep-th]}}.

\bibitem{Marino:2009jd}
M.~Mari\~no and P.~Putrov, ``{Exact Results in ABJM Theory from Topological
  Strings},'' \href{http://dx.doi.org/10.1007/JHEP06(2010)011}{{\em JHEP}
  {\bfseries 1006} (2010) 011},
\href{http://arxiv.org/abs/0912.3074}{{\ttfamily arXiv:0912.3074 [hep-th]}}.

\bibitem{Moens:2003JAC}
E.~Moens and J.~Van~der Jeugt, ``{A Determinantal Formula for Supersymmetric
  Schur Polynomials},'' \href{http://dx.doi.org/10.1023/A:1025048821756}{{\em
  J. Alg. Comb.} {\bfseries 17} (2003) 283--307}.

\bibitem{Hatsuda:2013yua}
Y.~Hatsuda, M.~Honda, S.~Moriyama, and K.~Okuyama, ``{ABJM Wilson Loops in
  Arbitrary Representations},''
  \href{http://dx.doi.org/10.1007/JHEP10(2013)168}{{\em JHEP} {\bfseries 1310}
  (2013) 168},
\href{http://arxiv.org/abs/1306.4297}{{\ttfamily arXiv:1306.4297 [hep-th]}}.

\bibitem{Dolivet:2006ii}
Y.~Dolivet and M.~Tierz, ``{Chern--Simons matrix models and Stieltjes--Wigert
  polynomials},'' \href{http://dx.doi.org/10.1063/1.2436734}{{\em J. Math.
  Phys.} {\bfseries 48} (2007) 023507},
\href{http://arxiv.org/abs/hep-th/0609167}{{\ttfamily arXiv:hep-th/0609167
  [hep-th]}}.

\bibitem{Rosso:1993JKTR}
M.~Rosso and V.~Jones, ``{On the invariants of torus knots derived from quantum
  groups},''
\href{http://dx.doi.org/10.1142/S0218216593000064}{{\em J. Knot Theory
  Ramifications} {\bfseries 2} (1993) 97--112}.

\bibitem{Kapustin:2010xq}
A.~Kapustin, B.~Willett, and I.~Yaakov, ``{Nonperturbative Tests of
  Three-Dimensional Dualities},''
  \href{http://dx.doi.org/10.1007/JHEP10(2010)013}{{\em JHEP} {\bfseries 1010}
  (2010) 013},
\href{http://arxiv.org/abs/1003.5694}{{\ttfamily arXiv:1003.5694 [hep-th]}}.

\bibitem{Aharony:2008gk}
O.~Aharony, O.~Bergman, and D.~L. Jafferis, ``{Fractional M2-branes},''
  \href{http://dx.doi.org/10.1088/1126-6708/2008/11/043}{{\em JHEP} {\bfseries
  0811} (2008) 043},
\href{http://arxiv.org/abs/0807.4924}{{\ttfamily arXiv:0807.4924 [hep-th]}}.

\bibitem{Basor:1994MN}
E.~L. Basor and P.~J. Forrester, ``{Formulas for the Evaluation of Toeplitz
  Determinants with Rational Generating Functions},''
  \href{http://dx.doi.org/10.1002/mana.19941700102}{{\em Math. Nachr.}
  {\bfseries 170} (1994) 5--18}.

\bibitem{Aganagic:2002wv}
M.~Aganagic, A.~Klemm, M.~Mari\~no, and C.~Vafa, ``{Matrix model as a mirror of
  Chern--Simons theory},''
  \href{http://dx.doi.org/10.1088/1126-6708/2004/02/010}{{\em JHEP} {\bfseries
  0402} (2004) 010},
\href{http://arxiv.org/abs/hep-th/0211098}{{\ttfamily hep-th/0211098}}.

\bibitem{Halmagyi:2003ze}
N.~Halmagyi and V.~Yasnov, ``{The spectral curve of the lens space matrix
  model},'' \href{http://dx.doi.org/10.1088/1126-6708/2009/11/104}{{\em JHEP}
  {\bfseries 0911} (2009) 104},
\href{http://arxiv.org/abs/hep-th/0311117}{{\ttfamily arXiv:hep-th/0311117}}.

\bibitem{Jockers:2012pz}
H.~Jockers, A.~Klemm, and M.~Soroush, ``{Torus Knots and the Topological
  Vertex},'' \href{http://dx.doi.org/10.1007/s11005-014-0687-0}{{\em Lett.
  Math. Phys.} {\bfseries 104} (2014) 953--989},
\href{http://arxiv.org/abs/1212.0321}{{\ttfamily arXiv:1212.0321 [hep-th]}}.

\bibitem{Stevan:2013tha}
S.~Stevan, ``{Torus Knots in Lens Spaces and Topological Strings},''
  \href{http://dx.doi.org/10.1007/s00023-014-0362-0}{{\em Ann. Henri
  Poincar{\'{e}}} {\bfseries 16} (2015) 1937--1967},
\href{http://arxiv.org/abs/1308.5509}{{\ttfamily arXiv:1308.5509 [hep-th]}}.

\bibitem{Borot:2013lpa}
G.~Borot, B.~Eynard, and N.~Orantin, ``{Abstract loop equations, topological
  recursion and new applications},''
  \href{http://dx.doi.org/10.4310/CNTP.2015.v9.n1.a2}{{\em Commun. Num. Theor.
  Phys.} {\bfseries 9} (2015) 51--187},
\href{http://arxiv.org/abs/1303.5808}{{\ttfamily arXiv:1303.5808 [math-ph]}}.

\bibitem{Borot:2013pda}
G.~Borot, A.~Guionnet, and K.~K. Kozlowski, ``{Large-$N$ asymptotic expansion
  for mean field models with Coulomb gas interaction},''
  \href{http://dx.doi.org/10.1093/imrn/rnu260}{{\em Int. Math. Res. Not.}
  {\bfseries 2015} (2015) 10451--10524},
\href{http://arxiv.org/abs/1312.6664}{{\ttfamily arXiv:1312.6664 [math-ph]}}.

\bibitem{Aganagic:2003db}
M.~Aganagic, A.~Klemm, M.~Marino, and C.~Vafa, ``{The Topological vertex},''
  \href{http://dx.doi.org/10.1007/s00220-004-1162-z}{{\em Commun. Math. Phys.}
  {\bfseries 254} (2005) 425--478},
\href{http://arxiv.org/abs/hep-th/0305132}{{\ttfamily arXiv:hep-th/0305132
  [hep-th]}}.

\bibitem{Klemm:2012ii}
A.~Klemm, M.~Mari\~no, M.~Schiereck, and M.~Soroush, ``{ABJM Wilson loops in
  the Fermi gas approach},''
  \href{http://dx.doi.org/10.5560/ZNA.2012-0118}{{\em Z. Naturforsch.}
  {\bfseries A68} (2013) 178--209},
\href{http://arxiv.org/abs/1207.0611}{{\ttfamily arXiv:1207.0611 [hep-th]}}.

\bibitem{Marino:2006hs}
M.~Mari\~no, ``{Open string amplitudes and large order behavior in topological
  string theory},'' \href{http://dx.doi.org/10.1088/1126-6708/2008/03/060}{{\em
  JHEP} {\bfseries 0803} (2008) 060},
\href{http://arxiv.org/abs/hep-th/0612127}{{\ttfamily arXiv:hep-th/0612127
  [hep-th]}}.

\bibitem{Bouchard:2007ys}
V.~Bouchard, A.~Klemm, M.~Mari\~no, and S.~Pasquetti, ``{Remodeling the
  B-model},'' \href{http://dx.doi.org/10.1007/s00220-008-0620-4}{{\em Commun.
  Math. Phys.} {\bfseries 287} (2009) 117--178},
\href{http://arxiv.org/abs/0709.1453}{{\ttfamily arXiv:0709.1453 [hep-th]}}.

\bibitem{Eynard:2012nj}
B.~Eynard and N.~Orantin, ``{Computation of open Gromov--Witten invariants for
  toric Calabi--Yau 3-folds by topological recursion, a proof of the BKMP
  conjecture},'' \href{http://dx.doi.org/10.1007/s00220-015-2361-5}{{\em
  Commun. Math. Phys.} {\bfseries 337} (2015) 483--567},
\href{http://arxiv.org/abs/1205.1103}{{\ttfamily arXiv:1205.1103 [math-ph]}}.

\bibitem{Aganagic:2000gs}
M.~Aganagic and C.~Vafa, ``{Mirror symmetry, D-branes and counting holomorphic
  discs},''
\href{http://arxiv.org/abs/hep-th/0012041}{{\ttfamily arXiv:hep-th/0012041
  [hep-th]}}.

\bibitem{Aganagic:2001nx}
M.~Aganagic, A.~Klemm, and C.~Vafa, ``{Disk instantons, mirror symmetry and the
  duality web},'' {\em Z. Naturforsch.} {\bfseries A57} (2002) 1--28,
\href{http://arxiv.org/abs/hep-th/0105045}{{\ttfamily arXiv:hep-th/0105045
  [hep-th]}}.

\bibitem{Aganagic:2012jb}
M.~Aganagic and C.~Vafa, ``{Large $N$ Duality, Mirror Symmetry, and a
  Q-deformed A-polynomial for Knots},''
\href{http://arxiv.org/abs/1204.4709}{{\ttfamily arXiv:1204.4709 [hep-th]}}.

\bibitem{Hatsuda:2013oxa}
Y.~Hatsuda, M.~Mari\~no, S.~Moriyama, and K.~Okuyama, ``{Non-perturbative
  effects and the refined topological string},''
  \href{http://dx.doi.org/10.1007/JHEP09(2014)168}{{\em JHEP} {\bfseries 1409}
  (2014) 168},
\href{http://arxiv.org/abs/1306.1734}{{\ttfamily arXiv:1306.1734 [hep-th]}}.

\bibitem{Eynard:2007nq}
B.~Eynard and N.~Orantin, ``{Topological expansion of mixed correlations in the
  Hermitian 2-matrix model and $x$--$y$ symmetry of the $F_g$ invariants},''
  \href{http://dx.doi.org/10.1088/1751-8113/41/1/015203}{{\em J. Phys.}
  {\bfseries A41} (2008) 015203},
\href{http://arxiv.org/abs/0705.0958}{{\ttfamily arXiv:0705.0958 [math-ph]}}.

\bibitem{Eynard:2013csa}
B.~Eynard and N.~Orantin, ``{About the $x$--$y$ symmetry of the $F_g$ algebraic
  invariants},''
\href{http://arxiv.org/abs/1311.4993}{{\ttfamily arXiv:1311.4993 [math-ph]}}.

\bibitem{Kimura:2014mua}
T.~Kimura, ``{Note on a duality of topological branes},''
  \href{http://dx.doi.org/10.1093/ptep/ptu141}{{\em PTEP} {\bfseries 2014}
  (2014) 103B04},
\href{http://arxiv.org/abs/1401.0956}{{\ttfamily arXiv:1401.0956 [hep-th]}}.

\bibitem{Kimura:2014vra}
T.~Kimura, ``{Duality and integrability of a supermatrix model with an external
  source},'' \href{http://dx.doi.org/10.1093/ptep/ptu163}{{\em PTEP} {\bfseries
  2014} (2014) 123A01},
\href{http://arxiv.org/abs/1410.0680}{{\ttfamily arXiv:1410.0680 [math-ph]}}.

\bibitem{Desrosiers:2009pz}
P.~Desrosiers and B.~Eynard, ``{Supermatrix models, loop equations, and
  duality},'' \href{http://dx.doi.org/10.1063/1.3430564}{{\em J. Math. Phys.}
  {\bfseries 51} (2010) 123304},
\href{http://arxiv.org/abs/0911.1762}{{\ttfamily arXiv:0911.1762 [math-ph]}}.

\bibitem{Brezin:2000CMP}
E.~Br\'ezin and S.~Hikami, ``{Characteristic Polynomials of Random Matrices},''
  \href{http://dx.doi.org/10.1007/s002200000256}{{\em Commun. Math. Phys.}
  {\bfseries 214} (2000) 111--135},
  \href{http://arxiv.org/abs/arXiv:math-ph/9910005}{{\ttfamily
  arXiv:math-ph/9910005}}.

\bibitem{HC:1957}
Harish-Chandra, ``{Differential Operators on a Semisimple Lie Algebra},''
  \href{http://dx.doi.org/10.2307/2372387}{{\em Amer. J. Math.} {\bfseries 79}
  (1957) 87--120}.

\bibitem{Itzykson:1979fi}
C.~Itzykson and J.-B. Zuber, ``{The Planar Approximation. II},''
\href{http://dx.doi.org/10.1063/1.524438}{{\em J. Math. Phys.} {\bfseries 21}
  (1980) 411--421}.

\bibitem{Tanaka:2012nr}
A.~Tanaka, ``{Comments on knotted 1/2 BPS Wilson loops},''
  \href{http://dx.doi.org/10.1007/JHEP07(2012)097}{{\em JHEP} {\bfseries 1207}
  (2012) 097},
\href{http://arxiv.org/abs/1204.5975}{{\ttfamily arXiv:1204.5975 [hep-th]}}.

\bibitem{Hama:2011ea}
N.~Hama, K.~Hosomichi, and S.~Lee, ``{SUSY Gauge Theories on Squashed
  Three-Spheres},'' \href{http://dx.doi.org/10.1007/JHEP05(2011)014}{{\em JHEP}
  {\bfseries 1105} (2011) 014},
\href{http://arxiv.org/abs/1102.4716}{{\ttfamily arXiv:1102.4716 [hep-th]}}.

\bibitem{Deguchi:1989zu}
T.~Deguchi and Y.~Akutsu, ``{Graded Solutions of the Yang--Baxter Relation and
  Link Polynomials},''
\href{http://dx.doi.org/10.1088/0305-4470/23/11/014}{{\em J. Phys.} {\bfseries
  A23} (1990) 1861--1876}.

\bibitem{Kauffman:1990ix}
L.~H. Kauffman and H.~Saleur, ``{Free fermions and the Alexander--Conway
  polynomial},''
\href{http://dx.doi.org/10.1007/BF02101508}{{\em Commun. Math. Phys.}
  {\bfseries 141} (1991) 293--327}.

\bibitem{Rozansky:1992td}
L.~Rozansky and H.~Saleur, ``{$S$ and $T$ matrices for the super U(1,1) WZW
  model: Application to surgery and three manifolds invariants based on the
  Alexander--Conway polynomial},''
  \href{http://dx.doi.org/10.1016/0550-3213(93)90326-K}{{\em Nucl. Phys.}
  {\bfseries B389} (1993) 365--423},
\href{http://arxiv.org/abs/hep-th/9203069}{{\ttfamily arXiv:hep-th/9203069
  [hep-th]}}.

\bibitem{Kashaev:1996kc}
R.~M. Kashaev, ``{The Hyperbolic volume of knots from quantum dilogarithm},''
  \href{http://dx.doi.org/10.1023/A:1007364912784}{{\em Lett. Math. Phys.}
  {\bfseries 39} (1997) 269--275},
\href{http://arxiv.org/abs/q-alg/9601025}{{\ttfamily arXiv:q-alg/9601025
  [math.QA]}}.

\bibitem{Murakami:2001AM}
H.~Murakami and J.~Murakami, ``{The colored Jones polynomials and the
  simplicial volume of a knot},''
  \href{http://dx.doi.org/10.1007/BF02392716}{{\em Acta Math.} {\bfseries 186}
  (2001) 85--104}, \href{http://arxiv.org/abs/math/9905075}{{\ttfamily
  arXiv:math/9905075 [math.GT]}}.

\bibitem{Gukov:2003na}
S.~Gukov, ``{Three-dimensional quantum gravity, Chern--Simons theory, and the
  A-polynomial},'' \href{http://dx.doi.org/10.1007/s00220-005-1312-y}{{\em
  Commun. Math. Phys.} {\bfseries 255} (2005) 577--627},
\href{http://arxiv.org/abs/hep-th/0306165}{{\ttfamily arXiv:hep-th/0306165
  [hep-th]}}.

\bibitem{Garoufalidis:2004GTM}
S.~Garoufalidis, ``{On the characteristic and deformation varieties of a
  knot},'' \href{http://dx.doi.org/10.2140/gtm.2004.7.291}{{\em Geom. Topol.
  Monogr.} {\bfseries 7} (2004) 291--309},
  \href{http://arxiv.org/abs/math/0306230}{{\ttfamily arXiv:math/0306230
  [math.GT]}}.

\bibitem{Khovanov:2008FM}
M.~Khovanov and L.~Rozansky, ``{Matrix factorizations and link homology},''
  \href{http://dx.doi.org/10.4064/fm199-1-1}{{\em Fund. Math.} {\bfseries 199}
  (2008) 1--91}, \href{http://arxiv.org/abs/math/0401268}{{\ttfamily
  arXiv:math/0401268 [math.QA]}}.

\bibitem{Khovanov:2008GT}
M.~Khovanov and L.~Rozansky, ``{Matrix factorizations and link homology II},''
  \href{http://dx.doi.org/10.2140/gt.2008.12.1387}{{\em Geom. Topol.}
  {\bfseries 12} (2008) 1387--1425},
  \href{http://arxiv.org/abs/math/0505056}{{\ttfamily arXiv:math/0505056
  [math.QA]}}.

\bibitem{Gukov:2004hz}
S.~Gukov, A.~S. Schwarz, and C.~Vafa, ``{Khovanov--Rozansky homology and
  topological strings},''
  \href{http://dx.doi.org/10.1007/s11005-005-0008-8}{{\em Lett. Math. Phys.}
  {\bfseries 74} (2005) 53--74},
\href{http://arxiv.org/abs/hep-th/0412243}{{\ttfamily arXiv:hep-th/0412243
  [hep-th]}}.

\bibitem{Dunfield:2005si}
N.~M. Dunfield, S.~Gukov, and J.~Rasmussen, ``{The Superpolynomial for Knot
  Homologies},'' \href{http://dx.doi.org/10.1080/10586458.2006.10128956}{{\em
  Experimental Math.} {\bfseries 15} (2006) 129--159},
\href{http://arxiv.org/abs/math/0505662}{{\ttfamily arXiv:math/0505662
  [math.GT]}}.

\end{thebibliography}\endgroup

\end{document}